\documentclass[12pt]{article}
\usepackage{latexsym}
\usepackage{amsmath,,calrsfs}
\usepackage{amsfonts}
\usepackage{amssymb}
\usepackage{amscd}
\usepackage{bbm}
\usepackage{fancybox}
\usepackage{cite}
\usepackage{amsmath,amsfonts,amsbsy}
\usepackage{pstricks,pst-node}
\usepackage[small,bf,hang]{caption2}
\usepackage{graphicx}
\usepackage{epsfig}
\usepackage{psfrag}
\usepackage{comment}

\usepackage{float}

\psset{unit=1.3cm,linewidth=.5pt,radius=.2}  % controls the size of diagrams

\usepackage{multirow}                     % tables cells spanning multiple rows
\usepackage{float}                          % to force floaters to stay Here
\usepackage{lscape}                         % landscape laid out pages
\usepackage{bm}

\newcommand{\bb}{\bar\beta}

\newcommand{\beq}{\begin{equation}}
\newcommand{\eeq}{\end{equation}}
\newcommand{\bi}{\begin{itemize}}
\newcommand{\ei}{\end{itemize}}
\newcommand{\bt}{\begin{tabular}}
\newcommand{\et}{\end{tabular}}
\newcommand{\bc}{\begin{center}}
\newcommand{\ec}{\end{center}}

\newcommand{\be}{\begin{equation}}
\newcommand{\ee}{\end{equation}}
\newcommand{\bea}{\begin{eqnarray}}
\newcommand{\eea}{\end{eqnarray}}
\newcommand{\ba}{\begin{array}}
\newcommand{\ea}{\end{array}}

\def\bbox{{\,\lower0.9pt\vbox{\hrule \hbox{\vrule height 0.2 cm
\hskip 0.2 cm \vrule height 0.2 cm}\hrule}\,}}
\newcommand{\dsl}{\pa \kern-0.5em /}

\font\mybb=msbm10 at 12pt
\def\bb#1{\hbox{\mybb#1}}
\def\bZ {\bb{Z}}
\def\bR {\bb{R}}
\def\bE {\bb{E}}
\def\bT {\bb{T}}
\def\bJ {\bb{J}}
\def\bM {\bb{M}}
\def\bN {\bb{N}}
\def\bL {\bb{L}}
\def\bG {\bb{G}}
\def\bH {\bb{H}}
\def\bC {\bb{C}}
\def\bO {\bb{O}}
\def\bA {\bb{A}}
\def\bX {\bb{X}}
\def\bP {\bb{P}}
\def\bQ {\bb{Q}}

\def\bV {\bb{V}}
\def\bK {\bb{K}}
\def\bS {\bb{S}}
\def\bY {\bb{Y}}
\def\bU {\bb{U}}
\def\bV {\bb{V}}
\def\bW {\bb{W}}
\def\bD {\bb{D}}
\def\bI{\bb{I}}

\def\bflam{\mbox{\boldmath $\lambda$}}

\def\bfpsi{\mbox{\boldmath $\psi$}}
\def\bfsigma{\mbox{\boldmath $\sigma$}}
\def\bfvartheta{\mbox{\boldmath $\vartheta$}}

\def\tr{{\rm tr}}
\makeatletter \@addtoreset{equation}{section} \makeatother

\def\slashchar#1{\setbox0=\hbox{$#1$}           % set a box for #1
   \dimen0=\wd0                                 % and get its size
   \setbox1=\hbox{/} \dimen1=\wd1               % get size of /
   \ifdim\dimen0>\dimen1                        % #1 is bigger
      \rlap{\hbox to \dimen0{\hfil/\hfil}}      % so center / in box
      #1                                        % and print #1
   \else                                        % / is bigger
      \rlap{\hbox to \dimen1{\hfil$#1$\hfil}}   % so center #1
      /                                         % and print /
   \fi}

\begin{document}

\begin{titlepage}%1
\begin{center}

\hfill  DAMTP-2017-35

\hfill  Imperial-TP-2017-ASA-01

\vskip 1.5cm

{\Large \bf  Twistor description of spinning particles in AdS}\\

\vskip 1cm

{\bf Alex S. Arvanitakis\,${}^{1,2,3}$, Alec E. Barns-Graham\,${}^{1}$ and 
Paul K.~Townsend\,${}^{1}$} \\

\vskip 1cm

{\em $^1$ \hskip -.1truecm
\em  Department of Applied Mathematics and Theoretical Physics,\\ Centre for Mathematical Sciences, University of Cambridge,\\
Wilberforce Road, Cambridge, CB3 0WA, U.K.\vskip 5pt }

\vskip .4truecm

{\em $^2$ \hskip -.1truecm
\em Department of Nuclear and Particle Physics,\\
Faculty of Physics, National and Kapodistrian University of Athens, \\
Athens 15784, Greece}

\vskip .4truecm

{\em $^3$ \hskip -.1truecm
\em The Blackett Laboratory,\\
Imperial College London, \\
Prince Consort Road London SW7 @AZ, U.K.\vskip 5pt}

\vskip 5pt
{email: {\tt A.Arvanitakis@imperial.ac.uk, P.K.Townsend@damtp.cam.ac.uk, \\A.E.Barnsgraham@damtp.cam.ac.uk}} \\

\end{center}

\vskip 0.5cm

\begin{center} {\bf ABSTRACT}\\[3ex]
\end{center}

The  two-twistor formulation of  particle mechanics in $D$-dimensional anti-de Sitter space for $D=4,5,7$, which linearises invariance 
under the  AdS isometry group $Sp(4;\bK)$ for $\bK=\bR,\bC,\bH$, is generalized to the massless $N$-extended  ``spinning particle''.   The twistor 
variables are gauge invariant with respect to the initial $N$ local worldline supersymmetries; this simplifies aspects of the quantum theory such as
implications of global gauge anomalies. We also give details of the two-supertwistor form of the superparticle, in particular the massive
superparticle on AdS$_5$.

\end{titlepage}

\newpage
\setcounter{page}{1} 
\tableofcontents

\newpage

%%%%%%%%%%%%%%%%%%%%%%%%%%%
%%%%%%%%%%%%%%%%%%%%%%%%%%%
\section{Introduction}

A general feature of particle, string or brane dynamics is that isometries of the background spacetime become symmetries of the 
particle, string or brane action. In the case of anti-de Sitter (AdS) backgrounds,  the AdS isometry group 
is usually realized non-linearly on the worldline, worldsheet or worldvolume fields, which complicates the extraction of physical consequences of these symmetries. This is especially
true for the supersymmetries of superparticles, superstrings or superbranes in the supersymmetric ``AdS$\times S$'' vacua of string/M-theory. 

There are various ways to linearize AdS (super)isometries. One, exploited in \cite{Claus:1998mw,Andrianopoli:1999kx},  is to regard AdS$_D$ as a hypersurface in 
$\bE^{2,d}$ for $d=D-1$.  An alternative is to exploit the fact that the AdS$_D$ isometry group is also  the conformal isometry  group of its $d$-dimensional Minkowski 
boundary;  twistor methods \cite{Penrose:1986ca,Atiyah:2017erd} are then available for some spacetime dimensions (as are supertwistor methods \cite{Ferber:1977qx,Shirafuji:1983zd}).  
This idea inspired a construction by  Claus et al. of an  action for a massive spin-zero particle in  AdS$_5$ for which the  AdS$_5$ isometries are  realized linearly on twistor variables \cite{Claus:1999zh}.

Twistor methods are available for $d=3,4,6,10$  (and supertwistors for $d=3,4,6$) \cite{Bengtsson:1987si,Cederwall:1993xe}. This is because these are the Minkowski space  dimensions for  which  the Lorentz group is $Sl(2;\bK)$ \cite{Kugo:1982bn,Sudbery,Manogue:1993ja,Viero} and the  conformal group is $Sp(4;\bK)$ \cite{Chung:1987in,Howe:1992bv}, where $\bK=\bR,\bC,\bH,\bO$ (the normed division algebras). This  fact suggests that the Claus et al. construction might be applicable more generally, and in particular to AdS$_{4,7}$ as well as AdS$_5$. The suggestion is attractive because the maximally supersymmetric ``AdS$\times S$'' vacua of string/M-theory have an AdS$_D$ factor precisely for $D=4,5,7$. 

A geometrical approach to this  problem was  formulated  by Cederwall  \cite{Cederwall:2000km}, who used the observation that an AdS$_D$ geodesic is the intersection of a plane in $\bE^{2,d}$ with the  AdS$_D$ hypersurface, the plane being specified by a  2-form on $\bE^{2,d}$. He then showed that a pair of twistor variables could be used to parametrize the 2-form  associated to a null or timelike geodesic of AdS$_5$, and a null geodesic of AdS$_{4,7}$, but more than two twistors would be needed for a timelike geodesic in AdS$_{4,7}$. The combined  results of \cite{Claus:1999zh,Cederwall:2000km} can be summarized as follows: a two-twistor formulation of the classical mechanics of a free spin-zero particle of mass $m$ in AdS$_D$ is possible for $D=5$, with a known action,  and it is also possible for $D=4,7$ but only if $m=0$.  

It is important to appreciate here that 
$m=0$ implies, and is implied by, a null worldline,  but it does {\it not} imply that the mass parameter $M$ of the quantum wave equation in AdS$_D$ is zero.  For $m=0$,  the standard zero-spin particle action is invariant under all {\it conformal isometries} of AdS$_D$ which implies, assuming preservation of conformal invariance upon quantization,   that the mass parameter of the quantum wave equation is $M=M_c$ with  $(M_cR/\hbar)^2= -D(D-2)/4$ \cite{Hawking:1973uf}. More generally, $M^2= M_c^2 + (m/\hbar)^2$, and the Breitenlohner-Freedman bound \cite{Breitenlohner:1982jf,Mezincescu:1984ev} is $(mR/\hbar)^2 \ge -1/4$, for which we gave a simple uncertainty-principle interpretation in \cite{Arvanitakis:2016vnp}.

The main result of  \cite{Arvanitakis:2016vnp} was a variant of the Claus et. al. construction  that applies uniformly to (super)particle mechanics in AdS$_D$ for $D=4,5,7$. It leads (in the ``bosonic'' case) to a manifestly $Sp(4;\bK)$-invariant action with canonical variables that are the entries of a  $4\times4$ matrix over $\bK$, which transforms  linearly with respect to both $Sp(4;\bK)$ and an $O(2;\bK)$ gauge group  (defined to preserve a $\bK$-hermitian quadratic form).  As the ${\bf 4}$ of $Sp(4;\bK)$ is a twistor for $d$-dimensional Minkowski space with $d= 2+ {\rm dim}\, \bK$, these canonical variables constitute  a ``two-twistor'';  i.e. a twistor doublet\footnote{In \cite{Arvanitakis:2016vnp} we called this a ``bi-twistor'', in accord with some earlier usage (e.g. \cite{Fedoruk:2004ru}), but as a ``bi-spinor'' is generally taken to mean a tensor represented as the sum of products of spinors we now prefer the terminology  ``two-twistor'', which is also in accord with some earlier usage (e.g \cite{deAzcarraga:2005ky}).}.

A peculiar feature of this construction is that it involves a mass-dependent change of variables that has the effect of eliminating the mass-dependence from the action,  which suggests that it is  actually valid only for $m=0$. This was verified in \cite{Arvanitakis:2016vnp} by a comparison of the  Noether charges for the manifest $Sp(4;\bK)$ symmetry with the Noether charges for invariance under the AdS isometries; they turn out to coincide only for $m=0$.  While this does not explain how the restriction to $m=0$ comes about (we postpone discussion of this point) it does confirm that the two-twistor action of \cite{Arvanitakis:2016vnp} indeed describes a particle (albeit massless) in AdS$_D$ for $D=4,5,7$. 

The  AdS$_5$ case is special because $O(2;\bC) \cong U(2)$  has  an `extra'  $U(1)$ factor unrelated to spin,  in contrast to  $O(2;\bK)\cong {\rm Spin}(1+ {\rm \dim}\,\bK)$ for  $\bK=\bR,\bH$. This (and the fact that $\bK=\bC$ for $D=5$) makes possible an $m$-dependent {\it complex} redefinition of  the twistor variables that `realigns'  the AdS isometry group $SU(2,2)_{AdS}$ with the manifest $Sp(4;\bC)\cong U(2,2)$ symmetry group; i.e. $U(2,2)\supset SU(2,2)_{AdS}$.  Its only other effect is to re-introduce the mass $m$ into the `extra' $U(1)\subset U(2,2)$, which coincides with the gauged $U(1)\subset U(2)$;  the results  of  \cite{Claus:1999zh} for the massive particle in AdS$_5$ are thereby recovered\footnote{They may also be recovered from the co-adjoint  orbit approach to particle dynamics \cite{Jiusi:2017xiy}; this approach was applied to massless particle dynamics in 
Mink$_d$ for $d=3,4,6$ in \cite{Howe:1992bv}.}.

In \cite{Arvanitakis:2016vnp} we extended these results to the superparticle \cite{Casalbuoni:1976tz,Brink:1981nb,Siegel:1983hh}.  We showed that `supersymmetrization' on 
Mink$_d$  `slices' of AdS$_D$ suffices because the resulting action has `hidden' supersymmetries. Here we confirm this explicitly for any $m$ 
when $D=5$ and for $m=0$ when
$D=4,7$. In  \cite{Arvanitakis:2016vnp} we identified the particular $\bK=\bR,\bC,\bH$ cases of relevance to the `AdS$\times$S' vacua of String/M-theory. Here we verify that  
the manifest $OSp({\cal N}|4;\bK)$ invariance supergroup coincides with the AdS isometry supergroup, again for any $m$ when $D=5$ and for $m=0$ when $D=4,7$. 

Even  in the exceptional AdS$_5$ case for which a (super)particle mass is compatible with its (super)twistor formulation, there is a quantum constraint on the mass coming from the possibility of a global $U(2)$ anomaly; we show that  the absence of this gauge anomaly requires the quantization condition $mR/\hbar\in \bZ$, where $R$ is the AdS$_5$ radius. This result depends on fact that $U(2)$ is a quotient of $U(1)\times SU(2)$ by $Z_2$; if the gauge group were $U(1)\times SU(2)$ then the quantization condition  would be $2mR/\hbar\in \bZ$. This weaker quantization condition is implicit in the results of  \cite{Claus:1999jj}, which makes use of earlier results in  \cite{Claus:1999xr}. 

The main purpose of this paper, however,  is to generalise the construction of \cite{Arvanitakis:2016vnp} to one that applies to  the $N$-extended
``spinning particle''  in AdS$_D$, again for $D=4,5,7$.  Like the superparticle, spin is incorporated via the addition of anticommuting worldline variables but now these are spacetime vectors and scalars rather than spinors, and they are introduced according to the requirement of {\it local worldline supersymmetry} rather than rigid spacetime supersymmetry. 

The original ``spinning particle''  action was a generalization of the standard action for a point particle in a Minkowski background to one  incorporating $N=1$  local worldline supersymmetry  \cite{Brink:1976sz}; it provides a  classical  (or pre-quantum) description of a  spin-$\tfrac{1}{2}$  particle in the sense that its quantization   yields the Dirac equation \cite{Brink:1976uf}.  The further generalization to  $N$-extended local worldline supersymmetry for $N>1$, and a local  $SO(N)$ gauge invariance, leads  to an action that describes  (at least for a four-dimensional Minkowski background)  a ``classical'' particle of spin $N/2$  \cite{Gershun:1979fb,Howe:1988ft,Howe:1989vn}.  

Here we are interested in an AdS background for the generic $N$-extended spinning particle.  For $N\le2$ the relevant action is just the specialization to AdS of the
action given in \cite{Howe:1988ft} for an arbitrary spacetime background, but it was erroneously claimed there that $N>2$ allows only flat backgrounds. This was corrected by 
Kuzenko and Yarevskaya \cite{Kuzenko:1995mg}, who showed that maximally symmetric background spacetime metrics are also allowed, in particular AdS; quantum aspects were subsequently explored by Bastianelli et al. \cite{Bastianelli:2014lia}.  The first task that we set ourselves in this paper is to summarize the status of the classical $N$-extended spinning particle in an AdS background using a formalism and notation  differing from  \cite{Kuzenko:1995mg} and \cite{Bastianelli:2014lia} but suited to our purposes. 

A crucial input to our subsequent construction of a  two-twistor action for a spinning particle  in AdS$_D$ for $D=4,5,7$  is the two-twistor action 
found in \cite{Mezincescu:2015apa} for a  massive  $N$-extended spinning particle in a $d$-dimensional Minkowski background for $d=3,4,6$.  A significant feature of that action (which carries over to the AdS$_D$ case)  is that  the  twistor variables, and the new anticommuting variables required  for non-zero spin, are all   {\it gauge invariant  with respect to the original local worldline supersymmetry}. The only remaining gauge-invariances other than time-reparametrization invariance are the local $SO(N)$ (for $N>1$) and those generated by the  $O(2;\bK)$ ``spin-shell'' constraints (which determine the Pauli-Lubanski 3-form \cite{Arvanitakis:2016wdn}).  Here we rederive these results using the  $Sl(2;\bK)$ and $Sp(4;\bK)$ notation to express them in a uniform way for $d=3,4,6$, and we take this opportunity to explain details of the new notation.  

With this Mink$_d$ result in hand, we proceed to  the AdS$_D$ case. As for the spin-zero particle, a comparison of the $Sp(4;\bK)$ Noether charges with the AdS$_D$ isometry charges shows that they coincide only for zero mass, so the two-twistor action must again be interpreted as describing a massless spinning particle in AdS$_D$.   Moreover, the complex redefinition that can be used in the AdS$_5$ case to circumvent this obstruction to 
non-zero mass for the zero-spin particle, and the (zero-superspin) superparticle,  no longer does so for the spinning particle. Because of this,  we limit our subsequent discussion of the quantum theory to the massless spinning particle. 

A general feature of $N$-extended spinning particle actions is that $N$ anticommuting variables become redundant in the $m\to0$ limit, in the sense that they are not required by the $N$-extended local supersymmetry  \cite{Mezincescu:2015apa}.  This is also a feature of our two-twistor action 
for the $N$-extended spinning particle in an AdS$_{4,5,7}$, as is to be expected from our conclusion that it describes a {\it massless} particle in these background spacetimes. Omission of the redundant anticommuting variables leaves us with a ``reduced'' two-twistor action and for $N=1$ we find that  
quantization yields results consistent with expectations derived from the standard action for the $N=1$ massless spinning particle. 

A complete discussion of the quantum theory might require something like a generalization of the results of \cite{Claus:1999jj}.  We leave this to the future.  Here we restrict 
ourselves, as we did for the superparticle in  \cite{Arvanitakis:2016vnp}, to an analysis of some  quantum implications of the classical anticommuting variables.  By omitting the redundant anticommuting variables we recover  expected  results for $D=4$ (a  massless particle of spin $N/2$) and for $D=5,7$ when $N=1$. The $D=5,7$ cases with $N>1$ are complicated by global $SO(N)$ anomalies; we postpone a  summary of our conclusions for these cases.  We close with a  discussion  of some issues  raised  by our results.

The $N=2$  spinning particle is special because of the possibility of including  an $SO(2)$ worldline Chern-Simons (WCS) term  \cite{Howe:1989vn}.
It turns out that the WCS term leads to a mismatch between the AdS isometry Noether charges and those of the manifest $Sp( 4;\bK)$ symmetry
of its would-be twistor formulation. The WCS term obstructs the twistor construction of the spinning particle in a way that is similar to the inclusion of non-zero mass for $D=5$ when $N>0$; the details are left to an Appendix.

%%%%%%%%%%%%%%%%%%%%%%%%%%%%%%
%%%%%%%%%%%%%%%%%%%%%%%%%%%%%
%%%%%%%%%%%%%%%%%%%%%%%%%%%%%%
\section{The $N$-extended spinning particle}\label{sec:Next}

We begin with a review of the status of the $N$-extended spinning particle, but in a notation that requires only the introduction of a background spacetime metric rather than 
a vielbein. We assume a spacetime (of unspecified dimension)  with metric $g_{mn}(x)$ in coordinates $x^m$.  The phase-space action for the $N$-extended spinning particle 
in this spacetime is a functional of maps from the particle's worldline  to a phase superspace with ``bosonic'' coordinates $\{x^m, p_m\}$ and $SO(N)$  $N$-plets of anticommuting 
(but Lorentz vector plus Lorentz scalar) coordinates  $\{\psi_i^m,\xi_i\}$ ($i=1,\dots,N$). The reparametrization invariant action takes the form 
\begin{equation}
S= \int \! dt \left\{ L_{\rm geom} + L_{\rm constraint}\right\}\, ,  
\end{equation}
for arbitrary worldline time $t$.  The ``geometrical'' part of the Lagrangian is 
\begin{equation}\label{LG}
L_{\rm geom} = \dot x^m p_m   + \frac{1}{2} \psi_i^m \dot \psi_i^n g_{mn} + \frac{1}{2} \xi_i \dot\xi_i \, , 
\end{equation}
where a sum over the index  $i$ is implicit, and we choose conventions for which the product of two `real' anticommuting variables is `real' {\it without} the customary additional imaginary unit factor. 
An  equivalent alternative expression is 
\begin{equation}
L_{\rm geom}= \dot x^m \pi_m  + \frac{1}{2} \psi_i^m (D_t\psi_i)^n g_{mn}(x) + \frac{1}{2} \xi_i \dot\xi_i \, , 
\end{equation}
where 
\begin{equation}
(D_t\psi_i)^p = \dot\psi_i^p + \dot x^m\Gamma_{mn}{}^p \psi_i^n \, , \qquad \Gamma_{mn}{}^p = \frac{1}{2} \left(g_{pm,n} + g_{pn,m}-g_{mn,p}\right)\, , 
\end{equation}
and 
\begin{equation}
\pi_m = p_m + \frac{1}{2} \Gamma_{mpq}\psi_i^p\psi_i^q\, , \qquad  \Gamma_{mpq} = \Gamma_{mp}{}^n g_{nq}\, . 
\end{equation}
The geometrical Lagrangian $L_{\rm geom}$ is the pullback to the worldline of a one-form on the phase superspace whose exterior derivative is the 
orthosymplectic 2-form 
\begin{eqnarray}\label{orthosN}
\Omega = dp_m dx^m + \frac{1}{2} d\psi_i^m d\psi_i^n g_{mn}- \frac{1}{2} dx^m d\psi_i^p g_{pq,m} \psi_i^q + d\xi_i d\xi_i\, . 
\end{eqnarray}
The inverse of $\Omega$ yields the canonical Poisson bracket (PB) relations:
\begin{equation}
\left\{ x^m ,p_n\right\}_{PB} = \delta^m_n\, , \quad \left\{\psi_i^m,\psi_i^n\right\}_{PB} = g^{mn}\delta_{ij} \, , \quad \left\{\xi_i,\xi_j\right\}_{PB} =\delta_{ij}\, , 
\end{equation}
and 
\begin{equation}
\left\{p_m,\psi_i^n\right\}_{PB} = \frac{1}{2}g^{nq}g_{qp,m}\psi_i^p\, , \qquad \left\{p_m,p_n\right\}_{PB} = - \frac{1}{4} g^{pq} g_{pr,m}g_{qs,n} \psi_i^r\psi_i^s\, . 
\end{equation}
These PB relations imply that 
\begin{equation}
\left\{ \pi_m ,\psi_i^n\right\}_{PB} = \Gamma_{mp}{}^n \psi_i^p \, , \qquad \left\{\pi_m,\pi_n\right\}_{PB} = \frac{1}{2} R_{mnrs} \psi_i^r\psi_i^s\, , 
\end{equation}
where $R_{mnrs}$ is the Riemann curvature tensor:
\begin{equation}
R^m{}_{nrs} =  2\partial_{[r}\Gamma_{s]n}{}^m + 2\Gamma_{p[r}{}^m \Gamma_{s]n}{}^p\, . 
\end{equation}

The constraint part of the Lagrangian is 
\begin{equation}
L_{\rm constraint} = -e{\cal H} -\chi_i {\cal Q}_i  - \frac{1}{2}f_{ij} {\cal J}_{ij}\, , 
\end{equation}
where $e,\chi_i, f_{ij}$ are Lagrange multiplers for the phase space constraints. For a particle of mass $m$,  
\begin{equation}\label{calJ}
{\cal Q}_i = \lambda_i^m\pi_m + m\xi_i\, , \qquad {\cal J}_{ij} = \psi_i^m \psi_j^n g_{mn}  +  \xi_i\xi_j\, . 
\end{equation}
We leave open for the moment the precise form of the Hamiltonian constraint function ${\cal H}$.  It must be chosen such that  the set of constraint functions 
is first-class, since they will then generate the gauge invariances of the action that are needed to allow the elimination of unphysical variables.

Using the PB relations given above one finds that 
\begin{equation}
\left\{{\cal J}_{ij},{\cal J}_{kl}\right\}_{PB} = 2\left(\delta_{k[i} {\cal J}_{j]l} - \delta_{l[j} {\cal J}_{i]k}\right)\, , 
\end{equation}
which shows that ${\cal J}_{ij}$ is a generator of $SO(N)$, and 
\begin{eqnarray}
\left\{{\cal J}_{ij} , {\cal Q}_k\right\}_{PB} &=& \left\{ \psi_i^p\psi_j^q g_{pq}, \psi_k^m \pi_m\right\}_{PB} + \mu\left\{\xi_i\xi_j,\xi_k\right\}_{PB} \nonumber \\
&=& 2 \delta_{k[j} \psi_{i]}^m\pi_m - 2 \psi_k^m \psi_i^p\psi_j^q\Gamma_{mqp} + \psi_k^m \psi_i^p\psi_j^q g_{pq,m} + 2\mu \delta_{k[j}\xi_{i]} \nonumber \\
&=& 2 \delta_{k[j} {\cal Q}_{i]}\, , 
\end{eqnarray}
which shows that the $N$ supercharges are the components of an $N$-vector of $SO(N)$.  One also finds that 
\begin{equation}\label{QQ}
\left\{{\cal Q}_i,{\cal Q}_j\right\}_{PB} = \left(g^{mn}\pi_m\pi_n + m^2\right) \delta_{ij} + \frac{1}{2} \psi_i^m\psi_j^n\psi_k^r\psi_k^s R_{mnrs}\, . 
\end{equation}
For $N=1$ the last term on the right-hand side  is zero, so 
\begin{equation}
\left\{{\cal Q},{\cal Q}\right\}_{PB} = 2{\cal H}\, , \qquad 2{\cal H}= g^{mn}\pi_m\pi_n + m^2\, ,   \qquad (N=1).
\end{equation}
In this case ${\cal H}$ is fixed by the requirement that the set of constraints is first class, and the algebra of constraint functions is then that of 
$N=1$ worldline supersymmetry. 

For $N=2$ we may use the  standard algebraic identities
\begin{equation}
R_{mnrs} \equiv R_{rsmn}\, , \qquad R_{m[nrs]} \equiv 0\, , 
\end{equation}
to deduce that 
\begin{equation}
\psi_i^m\psi_j^n \psi_k^r\psi_k^s R_{mnrs} \equiv   \frac{1}{2}\delta_{ij}  \psi_l^m\psi_l^n \psi_k^r\psi_k^s R_{mnrs}\, . 
\end{equation}
Using this identity, we can rewrite (\ref{QQ}) as 
\begin{equation}\label{QQ2}
\left\{{\cal Q}_i,{\cal Q}_j\right\}_{PB} = 2 \delta_{ij} {\cal H}\, ,
\end{equation}
where \cite{Howe:1988ft}
\begin{equation}\label{H2}
2{\cal H} = g^{mn}\pi_m\pi_n + m^2+ \frac{1}{4} \psi_l^m\psi_l^n \psi_k^r\psi_k^s R_{mnrs}\qquad (N=1,2) \, .
\end{equation}
This result also applies to $N=1$ because in that case the quartic fermion term is identically zero. 

For $N>2$ it may appear from (\ref{QQ})  that we must insist on a flat background in order to have a first-class set of constraints, but this is slightly 
too strong a condition; any maximally symmetric background  is also possible \cite{Kuzenko:1995mg}, as we now review for AdS.

%%%%%%%%%%%%%%%%%%%%%%%%%%%
%%%%%%%%%%%%%%%%%%%%%%%%%%
\subsection{$N>2$ and AdS}\label{sec:2.1}

For an AdS background, 
\begin{equation}\label{curv}
R_{mnrs} = - 2R^{-2} g_{m[r}g_{s]n}\, , 
\end{equation}
where $R$ is the constant AdS radius. In this case (\ref{QQ}) reduces to 
\begin{equation}\label{QQAdS}
\left\{{\cal Q}_i,{\cal Q}_j\right\}_{PB} = \left(g^{mn}\pi_m\pi_n + m^2\right) \delta_{ij} + R^{-2}J_{ik}J_{jk} \, ,
\end{equation}
where
\begin{equation}
J_{ij} =\psi_i^m \psi_j^n g_{mn} \, . 
\end{equation}
For $N=2$ we have the identity
\begin{equation}\label{scalarJ}
J_{ik}J_{jk} \equiv \delta_{ij} J^2 \, , \qquad \left(J^2 = \frac{1}{2} J_{ij}J_{ij}\right)\, . 
\end{equation}
This allows us to rewrite (\ref{QQAdS}) as 
\begin{equation}\label{QQAdS2}
\left\{{\cal Q}_i,{\cal Q}_j\right\}_{PB} =  2{\cal H} \delta_{ij} \, , \qquad 2{\cal H}= g^{mn}\pi_m\pi_n+ m^2 + R^{-2} J^2\, . 
\end{equation}
This is the specialization to AdS of (\ref{QQ2}) and  (\ref{H2}).  

For $N>2$ the $O(N)$ tensor $J_{ik}J_{jk}$ is not proportional to $\delta_{ij}$, so we must proceed differently. From the expression for ${\cal J}_{ij}$ in (\ref{calJ}) 
we see that 
\begin{equation}\label{JtoJ}
J_{ij} = {\cal J}_{ij} -\xi_i\xi_j\, , 
\end{equation}
which gives us 
\begin{equation}
J_{ik}J_{jk} =  {\cal J}_{ik}{\cal J}_{jk} + 2R^{-2} \xi_k \xi_{(i}{\cal J}_{j)k}\, ,
\end{equation}
and hence 
\begin{equation}\label{Jsquared}
J^2 = {\cal J}^2 - R^{-2}\xi_i\xi_j {\cal J}_{ij}\, , \qquad  \left({\cal J}^2 = \frac{1}{2} {\cal J}_{ij} {\cal J}_{ij}\right) \, .  
\end{equation}
We may now rewrite (\ref{QQAdS}) as 
\begin{equation}\label{QQa}
\left\{{\cal Q}_i,{\cal Q}_j\right\}_{PB} = 2{\cal H}(a) \, \delta_{ij} + K_{ij}(a)\, , 
\end{equation}
where,  for arbitrary  parameter $a$ (related to $b$ of \cite{Bastianelli:2014lia}), 
\begin{equation}\label{HamN}
2{\cal H}(a) = g^{mn}\pi_m\pi_n  + m^2 + aR^{-2}J^2 \, , 
\end{equation}
and
\begin{equation}
K_{ij}(a) = R^{-2}\left[ {\cal J}_{ik}{\cal J}_{jk} - a \delta_{ij} {\cal J}^2\right] 
- R^{-2}\left[2 \xi_{(i} {\cal J}_{j)k} -a \delta_{ij} \xi_l{\cal J}_{lk}\right]\xi_k \, . 
\end{equation}
This result is consistent with the requirement of first-class constraints for  any value of the constant $a$. 
We should find that ${\cal H}(a)$ is an $SO(N)$ singlet for any value of $a$, and we do because
\begin{equation}
\left\{{\cal J}_{ij},{\cal H}(a)\right\}_{PB} = \pi^m \left\{ \psi_i^p\psi_j^q g_{pq}, \pi_m\right\}_{PB} + \frac{a}{2R^2} \left\{ J_{ij}, J^2\right\}_{PB} =0\, . 
\end{equation}
The first term on the right hand side is zero because of a cancelation between the terms coming from the PB of $\pi_m$ with $g_{pq}$ and the
PB of $\pi_m$ with $\psi_i^p\psi_j^q$. The second, $a$-dependent, term is obviously zero. The only other PB of relevance is
\begin{equation}
\left\{{\cal Q}_i, {\cal H}(a)\right\}_{PB} =  \frac{(a-1)}{R^2} \left[ {\cal J}_{ij} \psi_j^m\pi_m - \xi_i\xi_j {\cal Q}_j\right]\, . 
\end{equation}
We see that the requirement of first-class constraints still allows arbitrary $a$, but the  $a=1$ case is special. 
For $a=1$ we have, for any $N$,  both
\begin{equation}
\left\{{\cal Q}_i, {\cal H}(1)\right\}_{PB}  =0 \, , 
\end{equation}
and 
\begin{equation}
\left\{{\cal Q}_i,{\cal Q}_j\right\}_{PB} = 2{\cal H}(1) \, \delta_{ij} + K_{ij}(1) \, , \qquad {\rm tr}\, K(1) =0\, . 
\end{equation}
For $N=2$ the traceless matrix $K(1)$ is zero and ${\cal H}(1)$ coincides with  the $N=2$ Hamiltonian constraint function 
of (\ref{QQAdS2}).

%%%%%%%%%%%%%%%%%%
\subsubsection{Killing vectors and Noether charges}

Every Killing vector field $k$ of the background spacetime corresponds to a symmetry of the spin-zero particle action, with 
Noether charge $k^mp_m$. How does this generalize to the spinning particle?
Using the PB relations given previously, one may verify that for any killing vector field $k$ the quantity
\begin{equation}\label{killingform}
K = k^m \pi_m - \frac{1}{2}\left(\partial_m k_n\right) \psi_i^m\psi_i^n
\end{equation}
 is a constant of the motion.  This shows that $K$ is the extension to the spinning particle of the Noether charge $k^mp_m$ of the spin-zero 
 particle associated to invariance  of the particle action under an isometry of the spacetime metric.  As such,  $K$ should be gauge invariant, which  requires
 \begin{equation}
 \left\{ K,{\cal Q}_i\right\}_{PB} =0\, . 
 \end{equation}
A PB calculation confirms that this condition is  satisfied.

%%%%%%%%%%%%%%%%%%%%%%%%%%%%%
%%%%%%%%%%%%%%%%%%%%%%%%%%%%%
%%%%%%%%%%%%%%%%%%%%%%%%%%%%%
\section{Minkowski spinning particle in  $Sl(2;\bK)$ notation}\label{sec:Mink}

A twistor action for the massive spinning particle in $d$-dimensional Minkowsi spacetime for $d=3,4,6$ was found in \cite{Mezincescu:2015apa}.
Our  aim in this section is to rederive that result using the $Sl(2;\bK)$ spinor notation of \cite{Arvanitakis:2016vnp}. This allows us to consider together the $d=3,4,6$ cases, and to
present further details of the notation.

Let  $\{x^\mu; \mu=0,1,\dots , d-1\}$ be Minkowski coordinates for a $d$-dimensional Minkowski space. The associated light-cone coordinates are 
\begin{equation}
x^\pm = x^0\pm x^{d-1}\, ,\qquad x = (x^1,\dots,x^{d-2})\, . 
\end{equation}
For $d=3,4,6,10$,  we may view the transverse position  in $\bR^{d-2}$ as an element of, respectively, $\bK=\bR,\bC,\bH,\bO$. These are the four 
normed division algebras over $\bR$; recall that for $x\in \bK$ the norm-squared of $x$ is $\bar x x$ where $\bar x$ is the $\bK$-conjugate of $x$.  
We may also represent a point in  Minkowski space by the  following $2\times2$ hermitian\footnote{Or ``$\bK$-hermitian in the terminology of  \cite{Arvanitakis:2016vnp}; we
henceforth simplify this to ``hermitian'', it being understood that that complex conjugation is generalized to conjugation with respect to $\bK$.} matrix
\begin{equation}
\bX = \left(\begin{array}{cc} x^+ & x \\ \bar x & x^- \end{array}\right) \, . 
\end{equation}
The  Lorentz group, with an element $\bL$,  acts on $\bX$ by 
\begin{equation}\label{Lorentz}
\bX \to \bL \bX \bL^\dagger\, , \qquad \det \left(\bL \bL^\dagger\right) =1\, .
\end{equation}
Let us examine this transformation separately for $\bK=\bR,\bC,\bH,\bO$.
\begin{itemize}

\item $\bK=\bR$. In this case the condition $\det(\bL\bL^\dagger)=1$ is equivalent to $|\det\bL|=1$. The subgroup for which $\det\bL=1$ is 
$Sl(2;\bR)$, the $d=3$ Lorentz group. 

\item $\bK=\bC$. In this case the condition on $\bL$ defines the group $Sl_2(2;\bC)$  whereas the Lorentz group 
is $Sl_1(2;\bC)$, for which $\det\bL=1$  (see e.g. \cite{Gilmore}); In other words, $Sl(2;\bC)\cong Sl_2(2;\bC)$, by
definition here, and the group action of (\ref{Lorentz})  therefore includes an  additional $U(1)$, which is such that 
the ``rotation'' subgroup  is $U(2)$  rather than $SU(2)$.

\item $\bK=\bH$. Although the determinant of an arbitrary quaternionic matrix cannot be  intrinsically  defined (i.e. without recourse to a matrix representation 
of the quaternion algebra) the determinant of an {\it hermitian} quaternionic matrix has an intrinsic definition\footnote{A useful reference
on determinants of quaternionic matrices is \cite{Aslaken}.},  which is such that $\bL \in Sl(2;\bH)\cong{\rm Spin}(1,5)$,  the $d=6$ Lorentz group, when 
$\det(\bL\bL^\dagger)=1$.

\item $\bK=\bO$.  In this case matrix multiplication becomes non-associative. In addition,
the number of real independent parameters of $\bL$ is now only $8\times4 -1 =31$, which is $14$ less than the 
$45$ required for ${\rm Spin}(1,9)$.  However, these two problems `cancel' because  the non-associativity of the octonions 
introduces another 14 real parameters, this being the dimension of its $G_2$ group of automorphisms; it is therefore possible 
to interpret $Sl(2;\bO)$ as the $d=10$ Lorentz group  \cite{Sudbery,Manogue:1993ja,Viero}.

\end{itemize}
In all these cases, a Lorentz vector is expressed as a bi-spinor,  a spinor being equivalent to a $2$-component $\bK$-valued column vector,
but in this paper we focus exclusively on the $\bK=\bR,\bC,\bH$ cases.

%%%%%%%%%%%%%%%%%%%%%%%%%%%%%
%%%%%%%%%%%%%%%%%%%%%%%%%%%%%
\subsection{The spin-zero particle and $Sl(2;\bK)$ spinors}\label{sec:Sl2K}

Consider now a free point particle of mass $m$ and zero spin in a 
$d$-dimensional  Minkowski spacetime, with metric $\eta = {\rm diag}(-1,1, \dots,1)$. The standard reparametrisation-invariant phase-space action is 
\begin{equation}\label{standact}
S= \int \! dt \left\{ \dot x^\mu p_\mu - \frac{1}{2} e\left(p^2+m^2\right)\right\},  \qquad p^2= \eta^{\mu\nu}p_\mu p_\nu\, , 
\end{equation}
where $p_\mu$ is the particle's $d$-momentum. For $d=3,4,6$ we may represent  $p$ by the $2\times2$ matrix
\begin{equation}
 \bP= \left(\begin{array}{cc} p_+ & p \\ \bar p & p_- \end{array}\right)\, , \qquad p_\pm = p_0 \pm p_1 \, , \qquad p\in \bK\, . 
 \end{equation}
 The Lorentz group acts on this matrix as follows
 \begin{equation}\label{co-Lorentz}
 \bP \to (\bL^\dagger)^{-1} \bP\,  \bL^{-1}\, . 
 \end{equation}
This transformation is such that 
\begin{equation}
\tr (\dot\bX \bP) \to \tr (\bL \dot\bX \bP\bL^{-1} )\, , 
\end{equation}
which is invariant for $\bK=\bR,\bC$ by the cyclic property of the trace. This property fails for $\bK=\bH$ but is still true for the real part of the 
trace (which we denote by $\tr_\mathbb{R}$).  In fact
\begin{equation}
\tr_\mathbb{R} (\dot\bX \bP) = 2 \dot x^m p_m\, . 
\end{equation}

Actually, we have not yet used the fact that $\bL$ represents a Lorentz transformation; we have only used the existence of $\bL^{-1}$ because no metric is needed for the contraction of $\dot x$ with $p$. Observe now that 
\begin{equation}
-p^2 \equiv  \det \bP \to \det \left[(\bL^\dagger)^{-1} \bP\,  \bL^{-1}\right] = \det\bP\, . 
\end{equation}
For $\bK=\bR,\bC$ the equality follows from the usual properties of determinants that allow us to rewrite the transformed determinant as $\det\bP/\det (\bL\bL^\dagger)$. 
This equality is not  obvious for $\bK=\bH$ but still true, as is easily verified explicitly for $2\times2$ matrices. 
We now see that the action (\ref{standact}) may be rewritten as 
\begin{equation}
S= \int\! dt \left\{\tfrac{1}{2}  \tr_\mathbb{R} (\dot\bX\bP) + \tfrac{1}{2} e \left(\det\bP -m^2\right)\right\}\, . 
\end{equation}

So far we have seen how to write the Lorentz scalars $\dot x^m p_m $ and $p^2$ in  $Sl(2;\bK)$ notation,  but any scalar product of
Lorentz vectors or co-vectors may be expressed in this notation. For any $2\times2$ hermitian matrix $\bX$, the Cayley-Hamilton theorem 
states that 
\begin{equation}\label{CH}
\bX^2 - \tr_\mathbb{R} (\bX) \, \bX +  (\det\bX) \, \bI_2  \equiv 0\, ,  
\end{equation}
which we may rewrite as \cite{Schray:1994fc}
\begin{equation}\label{CHprime}
\bX \tilde \bX \equiv - (\det\bX)\bI_2 = x^2 \bI_2\, , 
\end{equation}
where
\begin{equation}\label{xsquared}
\tilde \bX = \bX - (\tr_\mathbb{R} \bX)\bI_2 \qquad \left[\Leftrightarrow \ \bX = \tilde\bX - (\tr_\mathbb{R} \tilde \bX)\bI_2 \right]\, . 
\end{equation}
Taking the trace yields
\begin{equation}
\tr_\mathbb{R}(\bX\tilde\bX) = -2\det\bX = 2x^2\, , 
\end{equation}
and replacing $\bX$ by $\bX+\bY$ in this relation yields
\begin{equation}
2 x\cdot y = \tr_\mathbb{R}(\bX\tilde\bY)  = \tr_\mathbb{R} (\tilde\bX\bY)\, . 
\end{equation}
This tells us that if $x^m$ is represented by $\bX$ then $x_m$ is represented by $\tilde\bX$. In other words, in the $Sl(2;\bK)$ spinor formalism the raising and lowering of Lorentz vector indices  becomes the process of ``trace reversal''.

%%%%%%%%%%%%%%%%%%%%%%%%%%
%%%%%%%%%%%%%%%%%%%%%
\subsection{$N=1$ Spinning particle}

We now generalize to the $N=1$  massive spinning particle, in a $d$-dimensional Minkowski background. 
The standard reparametrisation-invariant phase-space action is 
\begin{equation}\label{standardspin}
S= \int\! dt \left\{ \dot x^\mu p_\mu + \tfrac{1}{2} \lambda^\mu \dot\lambda^\nu \eta_{\mu\nu} + \tfrac{1}{2} \xi\dot\xi
- \frac{1}{2}e\left(p^2+m^2\right) - \chi\left(\lambda^\mu p_\mu +m\xi\right)  \right\}\, .  
\end{equation}
For accord with the analysis to follow of the spinning particle in an AdS$_D$ background with $D=d+1$, we are now calling the anticommuting vector variable $\lambda^\mu$. 

In addition to representing $x^\mu$ and $p_\mu$ by the $2\times2$ hermitian matrices $\bX$ and $\bP$, we can represent $\lambda^\mu$ by the hermitian matrix 
\begin{equation}
\Lambda = \left(\begin{array}{cc} \lambda^+ & \lambda \\ \bar\lambda & \lambda^- \end{array}\right)\, , \qquad \lambda^\pm = \lambda^0\pm \lambda^1\, , 
\end{equation}
where $\lambda$ is anticommuting for $\bK=\bR,\bC$ and of the form $\lambda_0 + {\bf i} \cdot \bflam$ for $\bK=\bH$, where $\lambda_0$ and $\bflam$ are 
anticommuting (and ${\bf i}$ is the standard triplet of unit imaginary quaternions).  We shall use the term ``fermionic'' to cover all three cases. 

We may now rewrite the action (\ref{standardspin}) as
\begin{eqnarray}
S&=& \int\! dt \left\{ \tfrac{1}{2} \tr_\mathbb{R} (\dot\bX\bP)  + \tfrac{1}{4}\tr_\mathbb{R}(\tilde\Lambda \dot\Lambda) 
+\tfrac{1}{2}  \xi\dot\xi \right. \nonumber \\
&& \left. \ +\  \tfrac{1}{2} e \left(\det\bP -m^2\right) - \tfrac{1}{2}\chi \left[\tr_\mathbb{R}\left(\Lambda\bP\right)  +2 m\xi\right]\right\}\, . 
\end{eqnarray}
This action is invariant under the following worldline supertranslation gauge transformations:
\begin{equation}\label{lsusy}
\delta \bX = \alpha \tilde\bP + \epsilon\Lambda\, , \quad 
\delta\Lambda =  -\tilde\bP \epsilon\, , \quad  \delta\xi=-m\epsilon
\end{equation}
and 
\begin{equation}
\delta e= \dot\alpha + 2\epsilon\chi \, , \qquad \delta\chi=\dot\epsilon\, , 
\end{equation}
where $\alpha(t)$ is a commuting parameter and $\epsilon(t)$ an anticommuting parameter. 

%%%%%%%%%%%%%%%%%%%%%%%%%%
%%%%%%%%%%%%%%%%%%%%%%%
\subsection{Twistors and $O(2;\bK)$ gauge invariance}\label{subsec:O2}

The hermitian matrix $\bP$ may be written in the form
\begin{equation}\label{PtoU}
\bP =\mp \bU\bU^\dagger\, , 
\end{equation}
where the top sign applies if $p^0>0$ and the bottom sign applies if $p^0<0$. The
$2\times2$ matrix $\bU$ has the $Sl(2;\bK)$ transformation $\bU \to \bL \bU$. 
In other words, the two columns of $\bU$ are $Sl(2;\bK)$ spinors; they constitute a doublet of $O(2;\bK)$, which acts on 
$\bU$ from the right:
\begin{equation}
\bU \to \bU \bN \, , \qquad \bN \bN^\dagger = \bI_2 = \bN^\dagger\bN\, . 
\end{equation}
The expression for $\bP$ in terms of $\bU$ is therefore $O(2;\bK)$ invariant. This expression solves the mass-shell constraint 
$\det \bP=m^2$ provided that 
\begin{equation}\label{newms}
\det (\bU\bU^\dagger) = m^2\, , 
\end{equation}
which becomes a new mass-shell constraint. This shows that the matrix $\bU$ is invertible provided $m^2>0$. It has more independent 
components than $\bP$ but these will be unphysical provided the $O(2;\bK)$ transformations of $\bU$ are realized as a local symmetry
on the particle's worldline. In this case, $\bU(t)$ can be interpreted as a  choice of spatial frame at each point of the worldline; this is because
$O(2;\bK)$ is essentially the group of space rotations. Let us examine this assertion separately for $\bK=\bR,\bC,\bH,\bO$.

\begin{itemize}

\item $\bK=\bR$. In this case $\bN$ is an orthogonal $2\times2$ matrix and hence an element of $O(2)$, which contains the $SO(2)$ subgroup  of rotations 
in $\bE^2$.

\item $\bK=\bC$. In this case $\bN$ is a complex unitary $2\times2$  matrix and hence an element of $U(2)$. This contains the $SU(2)\cong {\rm Spin}(3)$ subgroup of rotations 
in $\bE^3$, but also an additional $U(1)$ factor (as anticipated in our earlier discussion of the Lorentz groups). 

\item $\bK=\bH$. In this case $\bN$ is a ``quaternionic unitary''  $2\times2$  matrix. Such matrices are elements of  the $Sp_2\cong USp(4) \cong {\rm Spin}(5)$ subgroup of 
rotations in $\bE^5$. 

\item $\bK=\bO$. We include this case only for completeness. The number of real parameters of $\bN$ is  now $8\times 4- 10= 22$, which is the right number to reduce the $32$ 
components of $\bU$ to the $10$ components of $\bP$. It is also  $14$ short of the $36$ needed for ${\rm Spin}(9)$, and we presume 
that $O(2;\bO)$ {\it could} be interpreted as ${\rm Spin}(9)$. However, generators of a  {\it gauge} invariance, when incorporated into a 
phase-space action need not form a Lie algebra; they need only be first class. 

\end{itemize}

Given that $\bU$ is invertible, we may introduce its inverse $\bV \equiv \bU^{-1}$ (the left and right inverses are equal even in the quaternionic case  \cite{Zhang}). Then 
\begin{equation}\label{tildePV}
\tilde\bP = \pm \det(\bU\bU^\dagger) \bV^\dagger\bV\,  \qquad \left(\bV = \bU^{-1}\right)\, , 
\end{equation}
which follows from a verification of the identity (\ref{CHprime}) with $\bX$ replaced by $\bP$; i.e. 
\begin{equation}
\bP\tilde\bP \equiv  - (\det\bP)\bI_2\, . 
\end{equation}
This expression for $\tilde\bP$  is a special case of a more general result, valid under the assumption that 
$\det(\bU\bU^\dagger)$ is non-zero. Given any hermitian $2\times2$ matrix $\bA$ transforming by conjugation under $O(2;\bK)$, we can construct 
from it the hermitian matrix
\begin{equation}
\mathbb{T}= \sqrt{\det(\bU\bU^\dagger)} \, \mathbb{V}^\dagger \bA\bV\, , 
\end{equation}
which has the transformation properties of a Lorentz vector. The hermitian matrix for the corresponding Lorentz covector is then 
\begin{equation}\label{tracereversalformula}
\tilde{\mathbb{T}}=\frac{1}{\sqrt{\det(\bU\bU^\dagger)} } \mathbb{U} \tilde\bA \mathbb{U}^\dagger\, . 
\end{equation}
This follows by verification of the identity (\ref{CHprime}) for $\bT$, assuming its validity for $\bA$:  
\begin{equation} 
\bT \tilde\bT = \bV^\dagger \bA \tilde\bA \bU^\dagger = - (\det \bA)\bI_2 = - (\det \bT)\bI_2\, .   
\end{equation}
The choice $\bA= \pm \sqrt{\det(\bU\bU^\dagger)}\, \bI_2$ yields $\bT= \tilde \bP$ and 
$\tilde \bT=\bP$. 

Now we solve the local supersymmetry constraint by writing $\Lambda$ ($\tilde\Lambda$)  in the form
\begin{equation}\label{lamsol}
 \Lambda = m\bV^\dagger \left(\Psi \pm \xi\bI_2 \right)\bV\, , \qquad \left(\tilde\Lambda = \frac{1}{m} \bU\left( \Psi\mp \xi \bI_2\right)\bU^\dagger\right)\, , 
 \end{equation}
 where
 \begin{equation}
 \Psi^\dagger= \Psi\, , \qquad \tr_\mathbb{R}\Psi =0\, . 
\end{equation}
As indicated, the fermionic  hermitian matrix $\Psi$ has zero real-trace. 
It is $Sl(2;\bK)$-inert but transforms by conjugation under the $O(2;\bK)$ gauge group:
\begin{equation}
\Psi \to \bN^\dagger \Psi \bN \, . 
\end{equation}
Substitution yields the Lagrangian
\begin{equation}
L = \frac{1}{4} \tr_\mathbb{R} \Psi\dot\Psi + \tr_\mathbb{R}(\dot \bU \bW^\dagger) \, , 
\end{equation}
where
\begin{equation}\label{incidence}
\bW= \pm \left[\bX \bU + \xi \bV^\dagger\Psi\right] + \frac{1}{2}\bV^\dagger \Psi^2\, .
\end{equation}
The fermionic nature of $\Psi$ implies that 
\begin{equation}\label{tracep2}
\tr_\mathbb{R}  \Psi^2 \equiv 0\, , 
\end{equation}
but $\Psi^2\not\equiv 0$.  For example, for $\bK=\bH$ we may write
\begin{equation}\label{psi1}
\Psi = \left(\begin{array}{cc} \varrho & \varsigma + {\bf i} \cdot \bfpsi \\ \varsigma - {\bf i}\cdot\bfpsi  & -\varrho\end{array}\right)\, , 
\end{equation}
for anticommuting singlets $\varrho$ and $\varsigma$ and anticommuting triplet $\bfpsi$, and then 
\begin{equation}\label{psi2}
\Psi^2 = \left(\begin{array}{cc} -{\bf i} \cdot (2\varsigma\bfpsi +\bfpsi \times\bfpsi)  &2\varrho (\varsigma + {\bf i} \cdot\bfpsi)  \\ -2 \varrho(\varsigma -{\bf i}\cdot\bfpsi)  & {\bf i} \cdot (2\varsigma\bfpsi - \bfpsi\times\bfpsi) \end{array}\right)\, . 
\end{equation}
Notice that $\Psi^2$ is {\it anti}-hermitian, since we are using a convention such that hermitian conjugation does not change the order of 
anticommuting factors. 

For  the $\bK=\bC$ case we have 
\begin{equation}
\Psi = \left(\begin{array}{cc} \varrho & \varsigma + i\psi \\ \varsigma - i\psi  & -\varrho\end{array}\right)\, , \quad 
\Psi^2 = \left(\begin{array}{cc} -2i \varsigma\psi & 2\varrho(\varsigma + i\psi) \\ 
-2\varrho(\varsigma - i\psi) & 2i\varsigma\psi\end{array}\right)\, , 
\end{equation}
and the $\bK=\bR$ case is found by setting $\psi=0$.  A special feature of these cases is that $\Psi^2$ is traceless. This statement is equivalent to (\ref{tracep2}) for $\bK=\bR$ 
but (\ref{tracep2}) leaves open the possibility of a non-zero imaginary part of the trace for $\bK=\bC,\bH$. Indeed, it {\it is} non-zero for $\bK=\bH$, but for $\bK=\bC$ it is zero.
This fact will become important in section \ref{sec:D=5}.

The ``incidence relation''  (\ref{incidence}) implies the identity
\begin{equation}\label{spin-shell}
\bG:=   \bU^\dagger \bW  - \bW^\dagger \bU - \Psi^2 \equiv 0\, . 
\end{equation}
This becomes a constraint in the action with independent phase space variables $(\bU,\bW,\Psi)$,  imposed by an anti-hermitian Lagrange multiplier $\bS$:
\begin{equation}\label{mtwist}
S= \int \! dt \left\{ \tr_\mathbb{R}(\dot\bU \bW^\dagger) + \frac{1}{4}\tr_\mathbb{R}(\Psi\dot\Psi)  - \tr_\mathbb{R}(\bS\bG) 
+ \frac{\ell}{2} \left(\det \bU\bU^\dagger -m^2\right)\right\}\, . 
\end{equation}
The new mass-shell constraint generates the new gauge-transformation
\begin{equation}\label{alpha-gauge}
\delta_\alpha \bW =  m^2\bV^\dagger \alpha \, , \qquad \delta_\alpha \ell = \dot\alpha +2\alpha\,  \tr_\mathbb{R}(\dot\bU\bV)\, , 
\end{equation}
for which invariance may be verified by means of  the identity
\begin{equation}
\frac{d}{dt}\left[\det(\bU\bU^\dagger)\right] \equiv 2 \det(\bU\bU^\dagger) \tr_\mathbb{R}(\dot\bU \bV)\, . 
\end{equation}
However, this gauge invariance is equivalent to time reparametrization invariance (for reasons explained in \cite{Routh:2015ifa} in the context of the
$d=6$ superparticle). 

Notice that $\xi$ has dropped out of the above action. This is because the new twistor variables are inert under the original worldline supersymmetry gauge transformations;  more precisely, $\bU$ is inert and $\bW$ is inert modulo a gauge transformation generated by the new mass-shell constraint, as we now explain.  Using the incidence relation  (\ref{incidence})  to compute the local supersymmetry transformation of $\bW$ from those of the initial 
variables $\bX$ and $\xi$, we find that 
\begin{equation}
\delta_\epsilon \bW = \pm \left[\delta_\epsilon\bX \bU + \delta_\epsilon\xi \bV^\dagger \Psi\right] = 
\pm \epsilon\left[\Lambda \bU - m\bV^\dagger \Psi\right]  = m \epsilon\xi \bV^\dagger\, , 
\end{equation}
where the last equality uses (\ref{lamsol}).   We see that $\bW$ is {\it not} strictly inert under the initial local worldline supersymmetry, but its transformation is  just an $\alpha$-gauge transformation of (\ref{alpha-gauge}) with parameter
\begin{equation}
\alpha = m^{-1}\epsilon \xi\, , 
\end{equation}
as originally found in \cite{Mezincescu:2015apa}.

%%%%%%%%%%%%%%%%%%%%%%%%%
%%%%%%%%%%%%%%%%%%%%%
\subsection{Twistors as $Sp(4;\bK)$ spinors}\label{subsec:Z}

Notice that 
\begin{equation}\label{twist-rewrite}
\tr_\mathbb{R}(\dot\bU \bW^\dagger) -  \frac{d}{dt}\left[ \frac{1}{2}\tr_\mathbb{R}\left(\bW \bU^\dagger\right)\right] = 
 \frac{1}{2} \tr_\mathbb{R}(\dot\bU \bW^\dagger - \dot\bW \bU^\dagger) \, . 
\end{equation}
Introducing the $4\times 2$ matrix 
\begin{equation}\label{deftwist}
\bZ = \left(\begin{array}{c} \bU \\ \bW \end{array}\right), 
\end{equation}
and the $4\times 4$ antisymmetric matrix
\begin{equation}
\Omega = \left(\begin{array}{cc} 0 & \bI_2 \\ -\bI_2 & 0 \end{array}\right)\, , 
\end{equation}
we may rewrite this as 
\begin{equation}
\tr_\mathbb{R}(\dot\bU \bW^\dagger)  = \tfrac{1}{2}\tr_\mathbb{R}(\Omega\, \dot\bZ \bZ^\dagger) + \frac{d}{dt}\left[\frac{1}{2}\tr_\mathbb{R}\left(\bW \bU^\dagger\right)\right] \, . 
\end{equation}
In this notation,  and omitting a total derivative, the action (\ref{mtwist}) becomes
\begin{equation}\label{mtwist2}
S= \int \! dt \left\{ \tfrac{1}{2}\tr_\mathbb{R}(\Omega\, \dot\bZ \bZ^\dagger)  + \tfrac{1}{4}\tr_\mathbb{R}(\Psi\dot\Psi) - \tr_\mathbb{R}(\bS\bG) 
-\frac{\ell}{2} \left(\det \bU\bU^\dagger -m^2\right)\right\}\, . 
\end{equation}
The spin-shell constraint functions $\bG$ may also be written in terms of $\bZ$:
\begin{equation}
\bG= -\bZ^\dagger \Omega \bZ - \Psi^2\, . 
\end{equation}

The advantage of rewriting the action in this way  is that it makes manifest the fact that only the  mass-shell constraint breaks what would otherwise be
an invariance under the following linear transformation of $\bZ$ with $4\times 4$ matrix parameter $\bM$ with entries in  $\bK$:
\begin{equation}
\bZ \to \bM \bZ \,,  \qquad \bM^\dagger \Omega \bM = \Omega\, . 
\end{equation}
This transformation defines the group that, following \cite{Sudbery},  we call $Sp(4;\bK)$. Let us examine the $\bK=\bR,\bC,\bH,\bO$ cases in turn. 

\begin{itemize}

\item $\bK=\bR$. In this case the $4\times4$ real matrix $\bM$ is an element of $Sp(4;\bR) \cong {\rm Spin}(2,3)$, the $d=3$ conformal group.

\item $\bK=\bC$. The $4\times4$  real antisymmetric matrix $\Omega$ is diagonalizable over $\bC$,  with doubly-degenerate 
eigenvalues $\pm i $, so $\bM$ is now an element of $U(2,2)$.  Ignoring discrete factors, this is equivalent to the product of $U(1)$ with 
$SU(2,2) \cong {\rm Spin}(2,4)$, the $d=4$ conformal group.  In other words, the group $Sl(4;\bC)$ is almost equivalent to the $d=4$ 
conformal group but, as for the rotation and Lorentz groups in four spacetime dimensions, it includes an additional $U(1)$ factor. 

\item $\bK=\bH$. In this case the $4\times4$ quaternionic matrix $\bM$ is an element of $Sp(4;\bH) \cong {\rm Spin}(2,6)$, the $d=6$ conformal group. 

\item $\bK=\bO$.  We comment on this case only for the sake of completeness. The $4\times4$ hermitian octonionic matrices that one might expect to 
span $Sp(4;\bO)$ have only $52$ real parameters, which is $14$ short of the $66$ needed for  the $d=10$ conformal group ${\rm Spin}(2,10)$. However 
a version of the ``add $14$ rule'' summarized earlier  is again applicable, so one may interpret  $Sp(4;\bO)$ as ${\rm Spin}(2,10)$ \cite{Chung:1987in}. 

\end{itemize} 

In summary, $Sp(4;\bK)$ is (essentially) the conformal group of $d$-dimensional Minkowski spacetime for $d=2+ {\rm dim}\, \bK$. A conformal group spinor is a twistor, which means that 
the  $4\times 2$ matrix  $\bZ$ is a ``two-twistor''; i.e. a twistor doublet acted upon from the left by $Sp(4;\bK)$ and from the right by the gauge group $O(2;\bK)$:
\begin{equation}
\bZ \to \bM \bZ \bN \, . 
\end{equation}
Returning to the action (\ref{mtwist}), we see that only the mass-shell constraint breaks the conformal invariance. This is apparently true even if we set $m^2=0$,  but in that case
the mass-shell constraint tells us that $\bU$ is no longer invertible, and this implies that there are additional gauge invariances, which  implies that the action is  no longer in canonical form (despite appearances).  One may expect  that when these additional gauge invariances are taken into account, the phase space 
action will be the standard one-twistor action for a massless particle in Mink$_d$ for $d=3,4,6$ with manifest  $Sp(4;\bK)$ invariance \cite{Bengtsson:1987si,Howe:1992bv}, as has been 
verified for the $d=3$ case in  \cite{Mezincescu:2015apa}.

%%%%%%%%%%%%%%%%%%%%%%%%%%
\subsubsection{$N>1$}

The extension to $N>1$ is almost immediate:  $\Psi$ becomes $\Psi_i$ with $i=1,\dots,N$, so the spin-shell constraint function is now
\begin{equation}
\bG= -\bZ^\dagger \Omega \bZ - \Psi_i\Psi_i\, , 
\end{equation}
and there is now an $SO(N)$ constraint with 
constraint function
\begin{equation}
{\cal J}_{ij} = \frac{1}{2} \tr_\mathbb{R}\left(\Psi_i\Psi_j\right)\, . 
\end{equation}
The action for the $N$-extended spinning particle of mass $m$ in a Mink$_d$ background is
\begin{eqnarray}\label{mtwistN}
S &=& \int \! dt \left\{ \tfrac{1}{2}\tr_\mathbb{R}(\Omega\, \dot\bZ \bZ^\dagger)  + \frac{1}{4}\tr_\mathbb{R}(\Psi_i\dot\Psi_i) - \tr_\mathbb{R}(\bS\bG) - \frac{1}{2} f_{ij} {\cal J}_{ij} \right.\nonumber \\
&& \qquad  \left. - \frac{\ell}{2} \left(\det \bU\bU^\dagger -m^2\right)\right\}\, . 
\end{eqnarray}
In the above formulae, a sum over repeated $SO(N)$ vector indices is implicit.

%%%%%%%%%%%%%%%%%%%%%%%%%%%%%%%
%%%%%%%%%%%%%%%%%%%%%%%%%%
%%%%%%%%%%%%%%%%%%%%%%%%%%
\section{Twistors and the  spinning particle in AdS$_D$}\label{sec:AdStwist}

We now return to the $N$-extended spinning particle in AdS$_D$.  Recall that the action takes the form 
\begin{equation}\label{ADSbackground}
S= \int\! dt\left\{ \dot x^m p_m + \frac{1}{2} \psi_i^m\dot\psi_i^n g_{mn} + \frac{1}{2} \xi_i \dot\xi_i - e{\cal H} - \chi_i{\cal Q}_i - \frac{1}{2}f_{ij} {\cal J}_{ij}\right\}\, . 
\end{equation}
There was some freedom in the choice of constraint functions, represented by the constant $a$. Choosing $a=0$ we have
\begin{equation}\label{ADScons}
{\cal H} = g^{mn}\pi_m\pi_n + m^2  \, , \qquad {\cal Q}_i = \psi_i^m \pi_m +m\xi_i\, .
\end{equation}

%%%%%%%%%%%%%%%%%%%%%%%%%%%%%%%
%%%%%%%%%%%%%%%%%%%%%%%%%%%%%%%
\subsection{Poincar\'e patch coordinates}

We shall now choose coordinates $x^m=\{x^\mu,z\}$  adapted to the foliation of AdS$_D$ by Minkowski hypersurfaces. The metric is 
\begin{equation}
ds^2 = \left(\frac{R}{z}\right)^2\left(dx^\mu dx^\nu \eta_{\mu\nu} + dz^2\right)\, . 
\end{equation}
The geometric part of the Lagrangian becomes
\begin{equation}\label{geom}
L_{\rm geom} = \dot x^\mu p_\mu +  \dot z p_z + \frac{1}{2}\eta_{\mu\nu} \lambda^\mu_i  \dot\lambda^\nu_i + \frac{1}{2} \zeta_i\dot\zeta_i  + \frac{1}{2} \xi_i\dot\xi_i\, , 
\end{equation}
where 
\begin{equation}
\lambda_i ^\mu = \left(\frac{R}{z}\right) \psi_i^\mu\, , \qquad \zeta_i = \left(\frac{R}{z}\right) \psi_i^z\, . 
\end{equation}

Now we turn to the constraints.  The non-zero components of the Levi-Civita affine connection are 
\begin{equation}
\Gamma_{zz}{}^z = - \frac{1}{z}\, , \quad \Gamma_{\mu\nu}{}^z = \frac{1}{z} \eta_{\mu\nu}\, , \qquad \Gamma_{\mu z}{}^\nu = \Gamma_{z\mu}{}^\nu = - \frac{1}{z} \delta_\mu^\nu\, . 
\end{equation}
Using this we find that 
\begin{equation} 
\pi_z = p_z\, , \qquad \pi_\mu = p_\mu + z^{-1} \lambda_\mu^i \zeta_i\, , 
\end{equation}
and using these relations we find that 
\begin{equation}\label{poincQ}
{\cal Q}_i =  \left(\frac{z}{R}\right)\left[p\cdot \lambda_i + \zeta_i p_z + \left(\frac{m R}{z}\right) \xi_i\right] + R^{-1} \lambda_i\cdot\lambda_j \zeta_j\, , 
\end{equation}
and that
\begin{equation}\label{Hzero}
2{\cal H} = \left(\frac{z}{R}\right)^2 \left[ p^2 + p_z^2 + \left(\frac{mR}{z}\right)^2\right]  + 2\frac{z}{R^2} p\cdot\lambda_i\zeta_i - R^{-2} \lambda_i\cdot\lambda_j \zeta_i\zeta_j \, , 
\end{equation}
where
\begin{equation}
p\cdot \lambda_i \equiv p_\mu \lambda_i^\mu\, , \qquad \lambda_i\cdot\lambda_j \equiv \lambda_i^\mu\lambda_j^\nu\eta_{\mu\nu}\, . 
\end{equation}

The $SO(N)$ constraint functions in the new variables are
\begin{equation}\label{Jcon}
{\cal J}_{ij} = \lambda_i\cdot \lambda_j + \zeta_i\zeta_j + \xi_i\xi_j \, , 
\end{equation}
and the constraint ${\cal J}_{ij}=0$ may be used to eliminate $\lambda_i\cdot \lambda_j$ in the expression for 
the supersymmetry constraint functions; the result for ${\cal Q}_i$ is
\begin{equation}\label{resultQ}
{\cal Q}_i =  \left(\frac{z}{R}\right)\left[ p\cdot\lambda_i + \zeta_i p_z + \left(\frac{mR+ \zeta_j\xi_j}{z}\right) \xi_i\right]\, . 
\end{equation}
We may now use both ${\cal J}_{ij}=0$  and  ${\cal Q}_i=0$ to simplify the expression for ${\cal H}$ to
\begin{equation}\label{resultH}
{\cal H} = \left(\frac{z}{R}\right)\left[ p^2 + p_z^2 + \left(\frac{mR+ \zeta_j\xi_j}{z}\right)^2\right]\, . 
\end{equation}
Finally, we can absorb the overall factors of $z/R$ in these expressions by a redefinition of the Lagrange multipliers, after which 
the action in the new variables becomes
\begin{equation}\label{actionPP}
S= \int dt \left\{ L_{\rm geom} - \frac{1}{2} \tilde e \, \tilde{\cal H} - \tilde{\chi_i} \, \tilde{\cal Q}_i - \frac{1}{2} f_{ij} {\cal J}_{ij}\right\}\, , 
\end{equation}
where 
\begin{equation}
\tilde{\cal H} =  p^2 + \Delta^2 \, , \qquad \tilde{\cal Q}_i =  p\cdot\lambda_i  + \Xi_i \, , 
\end{equation}
with 
\begin{equation}\label{defXi}
\Xi_i =  p_z\zeta_i  + \left(\frac{mR+ \zeta_j\xi_j}{z}\right) \xi_i\, , \qquad  \Delta^2 = p_z^2 + \left(\frac{mR+ \zeta_j\xi_j}{z}\right)^2\, .
\end{equation}

From the geometrical part of the Lagrangian, given by (\ref{geom}), we may read off  the Poisson brackets of the new canonical variables. The non-zero canonical Poisson brackets are
\begin{equation}\label{cPB1}
\left\{ x^\mu,p_\nu \right\}_{PB} = \delta_\nu^\mu \, , \qquad \left\{z,p_z\right\}_{PB} = 1\, , 
\end{equation}
and
\begin{equation}\label{cPB2}
\left\{\lambda_i^\mu,\lambda_j^\nu\right\}_{PB} = \eta^{\mu\nu}\delta_{ij} \, , \qquad \left\{\zeta_i,\zeta_j\right\}_{PB} = \delta_{ij} \, , \qquad \left\{\xi_i,\xi_j\right\}_{PB} =\delta_{ij}\, . 
\end{equation}
Using these relations  we find that 
\begin{equation}\label{PBXi}
\left\{\Xi_i,\Xi_j\right\}_{PB} = \Delta^2 \, \delta_{ij}\, , \qquad \left\{\Xi_i,\Delta^2\right\}_{PB} =0\, , 
\end{equation}
and hence that 
\begin{equation}
\left\{\tilde{\cal Q}_i, \tilde{\cal Q}_j\right\}_{PB} = 2 \tilde{\cal H}\,  \delta_{ij}\, , 
\end{equation}
which is the expected $N$-extended worldline supersymmetry algebra. 

%%%%%%%%%%%%%%%%%%%%%%%%%%%%%%%%
\subsubsection{The AdS isometries}

The Noether charges corresponding to the AdS isometries in the Poincar\'e patch coordinates are 
\begin{eqnarray}
P_\mu &=& p_\mu \nonumber \\
L^{\mu\nu} &=& 2x^{[\mu} p^{\nu]} - \lambda_i^\mu \lambda_i^\nu \, , \qquad D= x\cdot p + zp_z \nonumber \\
K^\mu &=& x^2p^\mu + z^2p^\mu - 2x^\mu(x\cdot p + zp_z) + 2\lambda_i^\mu (x\cdot \lambda_i + z\zeta_i)\, . 
\end{eqnarray}
In $Sl(2;\bK)$ bi-spinor notation 
\begin{eqnarray}\label{Ncharges}
P_\mu \ &\to& \  \bP \nonumber \\ 
L^\mu{}_\nu  + D\delta^\mu_\nu  \ &\to&  \ \bD \equiv \bP\bX + \frac{1}{2}\tilde\Lambda_i\Lambda_i + zp_z \bI_2 \nonumber \\
K^\mu \ &\to& \ \bK\, , 
\end{eqnarray}
where\footnote{We trust that this use of $\bK$ to denote the matrix of Noether charges associated to $K^\mu$ will not be confused with 
its use elsewhere to denote  one of the four division algebras $\bR,\bC,\bH,\bO$.}
\begin{equation}\label{bC}
\bK \equiv \frac{1}{2} \tilde\bP\left[\tr_\mathbb{R}(\bX\tilde\bX) + 2z^2\right] -\bX\left[\tr_\mathbb{R}(\bX\bP) + 2 zp_z\right] 
+ \Lambda_i\left[\tr_\mathbb{R}(\bX\tilde\Lambda_i)+ 2 z\zeta_i\right]\, . 
\end{equation}
Using the $Sl(2;\bK)$ matrix identities
\be\begin{aligned}\label{iden1}
\bX \bP\bX &\equiv \bX\,  \tr_\mathbb{R}(\bX\bP) - \frac{1}{2} \tr_\mathbb{R}(\bX\tilde\bX) \tilde\bP \, ,  \\
\Lambda_i\tilde\Lambda_i \bX - \bX\tilde\Lambda_i\Lambda_i &\equiv 2 \Lambda_i \tr_\mathbb{R}(\tilde\Lambda_i\bX) \, , 
\end{aligned}\ee
 we can write $\bK$ in the following alternative form:
\begin{equation}\label{bC2}
\bK=\bX\bP\bX + z^2\tilde\bP -2\bX zp_z
+ \frac12\left(\Lambda_i\tilde\Lambda_i \bX - \bX\tilde\Lambda_i\Lambda_i\right)+2 \Lambda_i z\zeta_i\, . 
\end{equation}
%%%%%%%%%%%%%%%%%%%%%%%%%%%%%%%
%%%%%%%%%%%%%%%%%%%%%%%%%%%%%%%%
\subsection{A change of anticommuting variables}

The first step in the passage to a two-twistor version of the action for a spinning particle in AdS$_D$ for $D=4,5,7$ is to make a 
redefinition of the  scalar\footnote{By ``scalar'' we mean here with respect to the $d$-dimensional Lorentz group.} anticommuting variables. 
First we define 
\begin{equation}\label{defZi}
Z_i = p_z \xi_i  - \left(\frac{mR + \zeta_j\xi_j}{z} \right)\zeta_i \, . 
\end{equation}
These phase-space functions satisfy  PB relations analogous to those of (\ref{PBXi}):
\begin{equation}
\left\{Z_i,Z_j\right\}_{PB} = \Delta^2 \, \delta_{ij}\  \, , \qquad \left\{Z_i, \Delta^2\right\}_{PB} =0\, . 
\end{equation}
In addition, 
\begin{equation}
\left\{Z_i,\Xi_j\right\}_{PB} =0\, .
\end{equation}
Next, we define the new variables
\begin{equation}
\xi_i' = \Xi_i/\Delta\, , \qquad \zeta_i' = Z_i/\Delta\, .
\end{equation}
These primed  variables satisfy the canonical PB relations 
\begin{equation}
\left\{\xi'_i,\xi'_j\right\}_{PB} = \delta_{ij} = \left\{\zeta'_i,\zeta'_j\right\}_{PB} \, , \qquad \left\{\xi'_i,\zeta'_j\right\}_{PB} =0\, . 
\end{equation}
The primed anticommuting variables are related to the unprimed ones by a rotation:
\begin{equation}
\left(\begin{array}{c} \xi'_i \\ \zeta'_i \end{array}\right) = 
\left(\begin{array}{cc} \cos\varphi &  \sin\varphi \\ -\sin\varphi & \cos\varphi \end{array}\right) \left(\begin{array}{c} \zeta_i \\ \xi_i \end{array}\right)\, , 
\end{equation}
where the angle $\varphi$ is such that
\begin{equation}
p_z = \Delta \cos\varphi\, , \qquad \frac{mR+\zeta_i\xi_i}{z} = \Delta\sin\varphi\, . 
\end{equation}

Now we make use of the following  {\bf Key Identity}: 
\begin{equation}\label{key-identity}
\boxed{\dot z p_z + \frac{1}{2}(\zeta_i\dot \zeta_i + \xi_i \dot \xi_i) \equiv  - zp_z \Delta^{-1}\dot\Delta + \frac{1}{2}(\zeta'_i\dot\zeta'_i + \xi'_i \dot\xi'_i) + \frac{d}{dt} (zp_z - mR\varphi)} 
\end{equation}
Provided that we can ignore the total derivative term, this identity allows us to view $\Delta$ as a canonical variable (with conjugate variable $zp_z/\Delta$). 
On AdS, in distinction to its universal cover, timelike geodesics are closed paths on which $\varphi$ increases by $2\pi$ on each traversal  \cite{Arvanitakis:2016vnp}. This means that 
the integral of $mR\dot\varphi$ is only defined modulo a  multiple of $2\pi mR$, which  suggests that the path integral will be well-defined for $m\ne0$ only if $mR\in \bZ$; we return to this issue in section \ref{sec:D=5}. 

We shall proceed on the assumption that the total derivative term on the right hand side of our ``key identity'' may be ignored. We may then use this identity
to rewrite the action (\ref{actionPP}) so that 
\begin{equation}\label{pretwist}
L_{\rm geom} = \dot x^\mu p_\mu - zp_z \Delta^{-1}\dot\Delta  + \frac{1}{2}\left( \lambda_i \cdot\dot\lambda_i + \zeta'_i\dot\zeta'_i + \xi'_i \dot\xi'_i \right) \, . 
\end{equation}
The constraint functions are now
\begin{equation}\label{tildecons}
\tilde{\cal Q}_i = p\cdot\lambda_i + \Delta\xi'_i\, , \qquad \tilde {\cal H} = \frac{1}{2} \left(p^2 + \Delta^2\right)\, , 
\end{equation}
where the  expression for $\Delta^2$ in terms
of the primed variables is 
\begin{equation}
\Delta^2= p_z^2 + \left(\frac{mR+ \xi'_i\zeta'_i}{z}\right)^2\, . 
\end{equation}
For $N>1$ we also have 
\begin{equation}
{\cal J}_{ij} = \lambda_i\cdot\lambda_j  + \xi'_i\xi'_j + \zeta'_i\zeta'_j\, . 
\end{equation}

%%%%%%%%%%%%%%%%%
\subsubsection{Conversion to $Sl(2;\bK)$ notation}

In  $Sl(2;\bK)$ matrix notation for the Lorentz $d$-vectors, eq. (\ref{pretwist}) becomes 
\begin{equation}\label{bitwist-geom}
L_{\rm geom} =  \tr_\mathbb{R}\left(\frac{1}{2} \dot{\mathbb X}\mathbb P + \frac{1}{4} \tilde\Lambda_i\dot\Lambda_i\right) 
+ \frac{1}{2} (\zeta'_i\dot\zeta'_i + \xi'_i \dot\xi'_i)- zp_z \Delta^{-1}\dot\Delta \, . 
\end{equation}
In addition, the constraints are now
\begin{equation}\label{bitwist-cons}
\tilde{\cal H} = \frac{1}{2} \left(-\det\mathbb P+\Delta^2\right) \, , \qquad \tilde{\cal Q}_i = \frac{1}{2} \tr_\mathbb{R}(\Lambda_i\mathbb P) +  \Delta \xi'_i
\end{equation}
and 
\begin{equation}\
{\cal J}_{ij} =  \frac{1}{2}\tr_\mathbb{R}\left(\tilde\Lambda_i\Lambda_j\right)+  \zeta'_i\zeta'_j +  \xi'_i \xi'_j\, . 
\end{equation}

If $\Delta$ were a constant, and if we could omit the $\zeta'_i$ variables, then the action would reduce to the action for a spinning particle 
of mass $\Delta$ in Mink$_d$ for $d=3,4,6$,  with $\xi_i'$ in place of $\xi_i$. This observation allows us to pass to a new two-twistor form 
of the action for the spinning particle in AdS$_D$ for $D=d+1$ by using the results of the previous section for the spinning particle 
in Mink$_d$. 

%%%%%%%%%%%%%%%%%%%%%%%%%%%%%%%
%%%%%%%%%%%%%%%%%%%%%%%%%%%%%
\subsection{Two-twistor action}\label{sec:bitwistact}

We now write $\mathbb P= \mp \mathbb{UU}^\dagger$ as we did for the particle in Minkowski space. The constraint $\det \mathbb P=\Delta^2$ becomes
\begin{equation}\label{DeltaUU}
\Delta^2=\det(\mathbb{UU}^\dagger)\, . 
\end{equation}
We write the constraint in this way because we no longer interpret it as a mass-shell constraint on $\bU$; instead, we interpret it as a constraint that determines
$\Delta$ in terms of $\bU$. Recalling the definition of $\tilde\bP$ in (\ref{tildePV}),  we now have
\begin{equation}
\tilde\bP = \pm \Delta^2 \bV^\dagger \bV\, . 
\end{equation}
In addition, it follows from (\ref{DeltaUU}) that 
\begin{equation}
\Delta^{-1} \dot \Delta=\tr_\mathbb{R}(\dot{\mathbb U}\mathbb{V})\,.
\end{equation}

We next solve the local supersymmetry constraints  by introducing $N$ traceless fermionic hermitian matrix variables $\Psi_i$ by
\begin{equation}
\label{Lambda_def}
\Lambda_i=\Delta \mathbb V^\dagger(\Psi_i\pm\xi'_i \mathbb I_2)\mathbb V\qquad \left[\Leftrightarrow \tilde \Lambda_i =\Delta^{-1}\mathbb{U}(\Psi_i\mp\xi'_i \mathbb I_2)\mathbb U^\dagger \right]\,,
\end{equation}
where $\Delta$ is now shorthand for $\sqrt{\det(\mathbb{UU}^\dagger)}$. Upon substitution for $\Delta$ and $\Lambda_i$,  the variable  $\xi'_i$ drops out, leaving us with the new lagrangian
\begin{equation}
L=\tr_\mathbb{R}\left(\dot{\mathbb{U}}\mathbb W^\dagger\right)  + \frac{1}{4}\tr_\mathbb{R}\left( \Psi_i \dot \Psi_i \right) + \frac{1}{2}\zeta'_i \dot\zeta'_i - \frac{1}{2}f_{ij} {\cal J}_{ij}\, , 
\end{equation}
where now
\begin{equation}
{\cal J}_{ij}= \frac{1}{2}\tr_\mathbb{R} \left(\Psi_i \Psi_j\right)+ \zeta'_i \zeta'_j\, , 
\end{equation}
and the variable $\mathbb{W}$  conjugate to $\mathbb{U}$ is found to be
\begin{equation}
\label{W}
\mathbb W=\pm(\mathbb X\mathbb U+\xi_i'\mathbb V^\dagger\Psi_i)+\frac{1}{2}\mathbb V^\dagger\Psi_i\Psi_i-zp_z\mathbb V^\dagger\,.
\end{equation}
This incidence relation implies the same identity as in the $d$-dimensional Minkowski case:
\begin{equation}\label{defbG}
\mathbb G :=\mathbb U^\dagger\mathbb W-\mathbb W^\dagger \mathbb U-\Psi_i\Psi_i\equiv 0\,.
\end{equation}
As before, we may view $\mathbb{W}$ as an independent canonical variable in the action by using a Lagrange multipler to impose $\mathbb{G}=0$ as a new phase-space constraint. 
This yields the action
\begin{equation}\label{action_spinning_ads_twistor}
S=\int dt\left \{ \tr_{\mathbb R}\left(\dot{\mathbb U} \mathbb W^\dagger\right) +\frac{1}{4}\tr_{\mathbb R}\left(\Psi_i\dot\Psi_i\right)+\frac{1}{2}\zeta'_i\dot\zeta'_i-\tr_{\mathbb R}\left(\bS\bG\right)- \frac{1}{2} f_{ij}\mathcal J_{ij}\right\}\, . 
\end{equation}
All constraints are first-class, with $\mathbb{G}$ generating an $O(2;\bK)$ gauge invariance.   As for the Minkowski case of section \ref{sec:Mink}, the absence of any fermionic constraints
implies that the two-twistor variables must be gauge invariant with respect to the initial $N$-extended local supersymmetries, and a calculation using the new incidence relation (\ref{W}) confirms this.

Let us now pause to consider how the action (\ref{action_spinning_ads_twistor}) differs from the action (\ref{mtwist}). One difference is that (\ref{action_spinning_ads_twistor}) involves $N$ additional 
anticommuting variables ($\zeta'_i$) that serve no obvious purpose, but we postpone discussion of this point. The most important difference is that  the mass-shell constraint  of (\ref{mtwist}) is absent from (\ref{action_spinning_ads_twistor}).  This has two immediate implications. One is that the phase space dimension has increased by $2$, which is consistent with the fact that we
now have a particle in a spacetime of dimension $D=d+1$. The second is that the action is now $Sp(4;\bK)$ invariant because the mass-shell constraint of  (\ref{mtwist}) is the only term in 
that action that is not $Sp(4;\bK)$ invariant. If we rewrite (\ref{action_spinning_ads_twistor})  in terms of the two-twistor $\bZ$ introduced in subsection \ref{subsec:Z} then we arrive at the
{\it manifestly} $Sp(4;\bK)$ invariant action 
\begin{eqnarray}\label{spinningact2}
S &=& \int \! dt \left\{ \tfrac{1}{2}\tr_\mathbb{R}\left(\Omega\dot\bZ \bZ^\dagger + \tfrac{1}{2}\Psi_i\dot\Psi_i\right)  + \tfrac{1}{2}\zeta'_i\dot\zeta'_i 
+ \tr_\mathbb{R}\left[\bS \left(\bZ^\dagger \Omega \bZ + \Psi_i\Psi_i\right)\right]  \right. \nonumber \\
&&\left.  \qquad  \quad  -\,  \tfrac{1}{2}f_{ij}\left[\tr_\mathbb{R} \left(\Psi_i\Psi_j \right) + 2\zeta_i'\zeta_j'\right]
\right\}\, . 
\end{eqnarray}
A peculiar feature of this action is that it is independent of the mass parameter $m$. This is due to the $m$-dependence of the change of  variables that we made  but  it is still puzzling: if there is no $m$-dependence in the action, how can it describe anything other than  a {\it massless} particle? 

We shall return to this question later, but a point to appreciate here is that there could  be more than one way to embed $Sp(4;\bK)$ into the full symmetry group of the action (which must be infinite-dimensional since any product of constants of motion is another constant of motion). There is, therefore, no guarantee that the linearly-realized $Sp(4;\bK)$ invariance group of the above action coincides with the $Sp(4;\bK)$ group of  AdS$_D$ isometries.  In fact, as we shall now explain, this correspondence holds only if $m=0$.

%%%%%%%%%%%%%%%%%%%%%%%%%%
%%%%%%%%%%%%%%%%%%%%%%%%%%%%%
\subsection{The $Sp(4;\bK)$  Noether charges}\label{sec:NC}

The Noether charges associated to the manifest $Sp(4;\bK)$ invariance of the action (\ref{spinningact2}) are  (passing over the one exception for 
$D=5$ that we return to later) the entries of the $4\times4$ matrix
\begin{equation}
\bJ  \equiv  \bZ \bZ^\dagger\, . 
\end{equation}
Following \cite{Arvanitakis:2016vnp}, we split $\bJ$ into its three independent $2\times 2$ blocks and evaluate them in spacetime using the incidence 
relations \eqref{Lambda_def} and \eqref{W}. 
First we have 
\begin{equation}
\mp\mathbb{UU}^\dagger = \mathbb P\,, \qquad \mathbb{UW}^\dagger= -\mathbb{P X} - z p_z \mathbb{I}_2 - \frac{1}{2} \tilde \Lambda_i \Lambda_i\, , 
\end{equation}
which coincide (passing over the $D=5$ exception alluded to above) with the Noether charges of  the Weyl subgroup of isometries of $d$-dimensional Minkowski spacetime. 
Then we have 
\begin{eqnarray}
\label{danger_noether_charge}
\pm\mathbb{WW}^\dagger &=&-\mathbb{XPX} -2 zp_z \mathbb{X}+ \left[z^2- \frac{(mR +\xi'_k\zeta'_k)^2}{\Delta^2}\right] \tilde{\mathbb{P}}
+\frac{1}{2} (\Lambda_i\tilde\Lambda_i \mathbb X -\mathbb{X} \tilde\Lambda_i \Lambda_i ) \nonumber\\
&& -\ \frac{1}{2} \frac{zp_z}{\Delta^2} (\Lambda_i\tilde\Lambda_i\tilde{\mathbb{P}} - \tilde{\mathbb{P}}\tilde\Lambda_i \Lambda_i) 
-\frac{1}{4\Delta^2} \Lambda_i\tilde\Lambda_i\tilde{\mathbb{P}}\tilde\Lambda_j \Lambda_j \, . 
\end{eqnarray}
We may simplify this expression by means of  the identities
\begin{eqnarray}
\Lambda_i\tilde\Lambda_i \tilde\bP - \tilde\bP\tilde\Lambda_i\Lambda_i &\equiv& 2 \Lambda_i\,  \tr_\mathbb{R}(\Lambda_i\bP) \,   \\
\Lambda_i\tilde\Lambda_i \tilde\bP \tilde\Lambda_j\Lambda_j &\equiv& \frac{1}{2}\tilde\bP\, \tr_\mathbb{R}(\Lambda_i\tilde\Lambda_j) \, \tr_\mathbb{R}(\Lambda_i\tilde\Lambda_j) 
- 2\Lambda_i \, \tr_\mathbb{R}(\Lambda_j\bP)\, \tr_\mathbb{R}(\Lambda_i\tilde\Lambda_j)\, .  \nonumber
\end{eqnarray}
Combining these two identities with the constraints ${\cal Q}_i=0$ and ${\cal J}_{ij}=0$ in the form
\begin{equation}
\tr_\mathbb{R}(\Lambda_i\bP) = -2\Delta\xi'_i \, , \qquad \tr_\mathbb{R}(\Lambda_i\tilde\Lambda_j) = -2(\zeta'_i\zeta'_j +\xi'_i\xi'_j)\, , 
\end{equation}
we deduce that
\begin{eqnarray}\label{iden2}
\Lambda_i\tilde\Lambda_i \tilde\bP - \tilde\bP\tilde\Lambda_i\Lambda_i &=&  -4\Delta \Lambda_i \xi'_i \, ,  \nonumber \\
\Lambda_i\tilde\Lambda_i \tilde\bP \tilde\Lambda_j\Lambda_j  &=& -4 (\xi'_i\zeta'_i)^2 \tilde\bP + 8\Delta (\Lambda_j\zeta'_j) (\xi'_i\zeta'_i)\, . 
\end{eqnarray}
Finally, using (\ref{iden2}) in (\ref{danger_noether_charge}) we deduce that
\begin{equation}\label{WWW}
\pm\bW\bW^\dagger =  \bK  - \left(\frac{mR}{\Delta}\right)^2 \tilde\bP  + 2\left(\frac{mR}{\Delta}\right) \left( \Lambda_i - \Delta^{-1}\xi'_i \right)\zeta'_i \, , 
\end{equation}
on the constraint surface,  where $\bK$ is the Hermitian matrix of (\ref{bC2}) that  represents  the AdS Noether charge $K^\mu$, but now 
expressed in terms of the new variables: 
\begin{eqnarray}\label{bC}
\bK &=&   \bX\bP\bX + z^2\tilde\bP -2\bX zp_z
+ \frac12\left(\Lambda_i\tilde\Lambda_i \bX - \bX\tilde\Lambda_i\Lambda_i\right)+2 \Lambda_i z\zeta_i\\
&& +  \Lambda_i\left[ 2 \left(\frac{zp_z}{\Delta}\right) \xi'_i -2 \left(\frac{mR + \xi'_k\zeta'_k}{\Delta}\right) \zeta'_i \right]\, . 
\end{eqnarray}
 The $m$-dependence of this expression  is purely the  result of the $m$-dependence of our change of variables. By means of the 
 formulae
 \begin{equation}\label{tildeP2}
\tilde\bP = \pm \Delta^2 \bV^\dagger \bV\, , \qquad \Lambda_i=  \Delta \bV^\dagger \left(\Psi_i \pm \xi'_i\, \bI_2\right)\bV\, , 
\end{equation}
we may further simplify the expression  for $\pm\bW\bW^\dagger$ to
\begin{equation}\label{WWW2}
\pm\bW\bW^\dagger =  \bK \mp \bV^\dagger \left[(mR)^2 \mp   2(mR)\Psi_i \zeta'_i \right] \bV\, . 
\end{equation}
This confirms, incidentally,  the invariance  of $\bW\bW^\dagger$ with respect to the original local worldline supersymmetries. However, it also shows that 
there is a discrepancy between the $Sp(4;\bK)$ Noether charges and the AdS$_D$ Noether charges {\it unless $m=0$}.

%%%%%%%%%%%%%%%%%%%%%%%%%%%
\subsubsection{Redundant anticommuting variables}

We observed above that the anticommuting variables $\zeta'_i$ of the two-twistor action (\ref{action_spinning_ads_twistor})  serve no obvious purpose. 
They are absent from the spin-shell constraints. For $N>1$ they appear in the $SO(N)$ constraint, but this is just because they form an $N$-vector of $SO(N)$. None of the essential features of the action (\ref{action_spinning_ads_twistor}) would change if  these variables  were absent; they are, in this sense, redundant. It appears that we could omit them but is there any other justification for doing so? 

There is, because we have just seen that (\ref{action_spinning_ads_twistor}) describes a {\it massless} particle in AdS$_D$ and inspection of the action (\ref{ADSbackground}) from which we started shows that the variables $\xi_i$ of that  action are similarly redundant when $m=0$ (in fact, for  the $N=1$ case,  $\zeta'=\xi$ when $m=0$).  The massive spinning particle in a Minkowski background also has  anticommuting variables that become redundant in a massless limit  \cite{Mezincescu:2015apa} and, unless omitted,  they lead to a reducible space of polarisation states;  we should expect the same to be true for an AdS background. 

Omitting the variables $\zeta'_i$ is equivalent to imposing $\zeta'_i=0$ as  additional, but second-class, constraints.  We could implement this in the 
two-twistor action  (\ref{action_spinning_ads_twistor}) by means of  additional Lagrange multipliers but it is obviously simpler  to directly set $\zeta'_i=0$ to get
\begin{equation}\label{simpleract}
S = \int \! dt \left\{ \tfrac{1}{2}\tr_\mathbb{R}\left[\Omega\dot\bZ \bZ^\dagger + \tfrac{1}{2}\Psi_i\dot\Psi_i + \bS \left(\bZ^\dagger \Omega \bZ + \Psi_i\Psi_i\right)\right] - \tfrac{1}{2}f_{ij}\tr_\mathbb{R} \left(\Psi_i\Psi_j \right) 
\right\}\, . 
\end{equation}
When the $\zeta'_i$ variables are similarly omitted from  (\ref{WWW2}), this formula
simplifies to 
\begin{equation}\label{WWW3}
\pm\bW\bW^\dagger =  \bK \mp (mR)^2 \bV^\dagger\bV\, . 
\end{equation}

%%%%%%%%%%%%%%%%%%%%%%%%%
%%%%%%%%%%%%%%%%%%%%%%%%
%%%%%%%%%%%%%%%%%%%%%%
\section{Non-zero mass for AdS$_5$}\label{sec:D=5}

For the AdS$_5$ case the matrices $\bU$ and $\bW$ are complex, rather than real or quaternionic.  In this case, we may replace $\bW$ 
by a new independent complex matrix variable $\breve\bW$ by setting
\begin{equation}\label{redef}
\bW = \breve\bW+ imR\,  \bV^\dagger\,  .
\end{equation}
Substitution yields
\begin{eqnarray}
\tr_\mathbb{R}(\dot \bU\bW^\dagger) &=&  \tr_\mathbb{R}(\dot \bU\breve\bW^\dagger) + mR\,  \tr_\mathbb{R}(i\dot \bU\bV) \nonumber \\
&=& \tr_\mathbb{R}(\dot \bU\breve\bW^\dagger) +  \frac{d}{dt} \left[(mR)\arg(\det \bU^\dagger)\right]\, ,
\end{eqnarray}
so the geometric part of the Lagrangian  is unchanged if we discard the total derivative.

In fact, the entire action (\ref{spinningact2})  has the same form in terms of $\breve\bW$ as it did in terms of $\bW$, {\it except for the trace of the spin-shell constraint}. To see this we observe that 
\begin{equation}\label{UWchange}
 \bU\bW^\dagger = \bU\breve\bW^\dagger- imR\, \bI_2\, , 
 \end{equation}
from which it follows that
\begin{equation}
\bG= \bU^\dagger \breve\bW- \breve\bW^\dagger\bU - \Psi_i\Psi_i  -2imR \, \bI_2  \, . 
\end{equation}
The parameter $mR$ contributes only to the trace of $\bG$, which is the `extra'  $U(1)$ part.  

The $Sp(4;\bC)\cong U(2,2)$ invariance of this new action 
in terms of $\bU$ and $\breve\bW$ may be made manifest by writing it in terms of the new twistor variables
\begin{equation}
\breve\bZ = \left(\begin{array}{c} \bU \\ \breve\bW\end{array}\right)\, . 
\end{equation}
The result, if we omit the redundant anticommuting variables $\zeta'_i$,  is the action 
\begin{equation}\label{simpleractm}
S = \int \! dt \left\{ \tfrac{1}{2}\tr_\mathbb{R}\left[\Omega\dot{\breve\bZ} \breve\bZ^\dagger + \tfrac{1}{2}\Psi_i\dot\Psi_i -
\bS \bG\right]- \tfrac{1}{2}f_{ij}\tr_\mathbb{R} \left(\Psi_i\Psi_j \right) \right\}\, , 
\end{equation}
where
\begin{equation}
\bG =  - \breve\bZ^\dagger \Omega \breve\bZ - 2imR - \Psi_i\Psi_i\, . 
\end{equation}
The Noether charges implied by the manifest $Sp(4;\bC)\cong U(2,2)$ symmetry are now contained in the $4\times 4$ hermitian matrix
\begin{equation}
\breve\bJ= \breve\bZ \breve\bZ^\dagger\, . 
\end{equation}
The action (\ref{simpleractm}) is formally the same as (\ref{simpleract}) except that the mass now appears in the  $U(1)$ constraint  imposed by the trace of the antihermitian Lagrange multiplier $\bS$.  For $N=0$, it is precisely the action for a spin-zero particle of mass $m$ in AdS$_5$  of \cite{Claus:1999zh}.   For $N>0$ we appear to have a  ``spinning''  extension of this massive particle action, but we have still to check whether the Noether charges $\breve\bJ$ are those implied by invariance under AdS$_5$ isometries.

Strictly speaking, what we have to check is that the  AdS$_5$ Noether charges are those combinations of the components of $\tilde\bJ$ that generate  the 
$SU(2,2)$ subgroup of $U(2,2)$, so the status  of the `extra'  $U(1)$ factor  requires clarification. Its generator is  the imaginary 
part of the trace of $\bU\bW^\dagger$ or, equivalently,  the real trace of $i\bU\bW^\dagger$. However, 
\begin{equation}\label{extragen}
 \tr_\mathbb{R}(i\bU\bW^\dagger) = - \frac{i}{2} \tr \left(\bU^\dagger\bW - \bW^\dagger\bU\right) =  - \frac{i}{2} \tr \, \bG\, . 
 \end{equation}
The last of these equalities  relies on the fact that $\Psi_i\Psi_i$ has zero trace for $\bK=\bC$ (and not merely zero real trace); this was noted for $N=1$ in section \ref{subsec:O2} but the result extends immediately to $N>1$.   What this equality shows is that  not all components of  $\bJ$ are  Noether charges (for $D=5$) because one combination is the constraint function for the  `extra'  $U(1)$ factor in the  $U(2)$ gauge group\footnote{For presentational simplicity we ignore the distinction between $U(2)$ and $U(1)\times SU(2)$ here, although it will become important below.}, and the same is true of $\breve\bJ$.

We are now in a position to return to the problem of the $(mR)^2$ term in the expression (\ref{WWW2}) for $\pm \bW\bW^\dagger$. The new Noether charges are
\begin{eqnarray}\label{checkW2}
\pm \breve\bW\breve\bW^\dagger&=& \pm \bW\bW^\dagger \pm imR\left(\bW\bV - \bV^\dagger\bW^\dagger\right) \pm (mR)^2\bV^\dagger\bV\nonumber \\
&=& \bK \mp i(mR) \, \bV^\dagger \left(\bU^\dagger\bW -\bW^\dagger\bU\right)\bV \nonumber \\
&=& \bK \mp i(mR) \, \bV^\dagger (\Psi_i\Psi_i)\bV\, , 
\end{eqnarray}
where the second line uses (\ref{WWW3}) and the last line uses the spin-shell constraint $\bG=0$.  
For the spin-zero  ($N=0$) particle the last term is absent, so there is no longer a discrepancy  between the $SU(2,2)$ Noether charges  
and the AdS$_5$ isometry Noether charges. 

For $N\ge1$ we have merely replaced the original discrepancy for $m\ne0$ by another one
that cannot be eliminated in a similar way.  The `discrepancy'  is itself a conserved charge having the same PB relations with the other 
AdS isometry charges as does $\bK$, except that it has zero PB with $\bP$ and hence no effect on the algebra. In other words, the conserved 
charges $\bP, \bD$ and 
\begin{equation}\label{discrep}
\breve\bK = \bK \mp i(mR) \, \bV^\dagger (\Psi_i\Psi_i)\bV
\end{equation}
span an algebra that is isomorphic to the AdS isometry algebra, but it is not the AdS isometry algebra because the 
extra term in $\breve\bK$ is 
\begin{equation}
\mp i(mR) \bV^\dagger (\Psi_i\Psi_i)\bV = - (i/\Delta^2)\left[ \bP \tilde\Lambda_i\Lambda_i +2\Delta\xi'_i \Lambda_i\right] \, ,  
\end{equation}
which is  incompatible with the general  form (\ref{killingform}) because of the inverse $\Delta$ factors.

%%%%%%%%%%%%%%%%%%%%%%%%%%%%%%%
\subsection{The $U(2)$ gauge anomaly and mass quantization}

We now consider some implications of the quantum theory for the Claus et al. action \cite{Claus:1999zh}, which in our notation is 
the $N=0$ case of (\ref{simpleractm}):
\begin{equation}
S = \int \! dt \left\{ \frac{1}{2}\tr_\mathbb{R}\left[\Omega\dot{\breve\bZ} \breve\bZ^\dagger  + \bS\left(\breve\bZ^\dagger \Omega \breve\bZ + 2imR\right)\right] \right\}\, . 
\end{equation}
We may rewrite this as $S=S_0+ S_m$, where $S_0$ is the action for $m=0$ and 
\begin{equation}
S_m = mR \int\! dt\,  \tr_\mathbb{R}(i\bS)\, . 
\end{equation}
This term is essentially a worldline Chern-Simons term for the $U(1)$ gauge group contained in $U(2)$.  The qualification ``essentially'' could have been omitted if the gauge group were $U(1)\times SU(2)$ because then $S_m$ would be a WCS term for the $U(1)$ factor and we could ignore the $SU(2)$ factor. However, it is important for the quantum theory that \begin{equation}
U(2) = [U(1)\times SU(2)]/\bZ_2\, , 
\end{equation}
because the quotient by $\bZ_2$  makes a difference.
The (finite) $U(2)$ gauge transformations are 
\begin{equation}
\breve\bZ \to \breve\bZ G \, , \qquad \bS \to G^{-1} \bS G + G^{-1}\dot G\, ,
\end{equation}
where the parameter $G(t)$ is a map from the worldline to the $U(2)$ gauge group. The action is invariant if $m=0$, but for $m\ne0$ we have
\begin{equation}\label{deltaSm}
S_m \to S_m + mR\int \! dt\, \tr_\mathbb{R}(i\dot GG^{-1})\, . 
\end{equation}
In the context of the Euclidean path integral\footnote{We may pass over  the question 
of how the analytic continuation from  Lorentzian to Euclidean spacetime metric is accomplished in the two-twistor formulation because the mass term in the action 
is independent of the spacetime metric.} we must consider closed worldlines, in which case the maps $G(t)$ may have a non-zero winding number, specified by an integer since 
$\pi_1\left(U(2)\right) =\bZ$.  For example, if we make the identification $t\sim t +1$ then representative maps in these integer homotopy classes are 
\begin{equation}
G_n (t) = \left[\exp \left\{-i\pi \left(\bI_2 + \sigma_3\right)t\right\}\right]^n \qquad n\in \bZ \, . 
\end{equation}
This implies $\tr_\mathbb{R}(i\dot GG^{-1})=2n\pi$, and using this in  (\ref{deltaSm}) we deduce that 
\begin{equation}
e^{iS_m} \to  \left[e^{(mR) 2\pi i}\right]^n e^{iS_m}\, . 
\end{equation}
It follows that $U(2)$ invariance of the path integral requires the quantization condition 
\begin{equation}
mR \in \bZ\, . 
\end{equation}
We must remind the reader here that the mass $m$ in the classical action is not the same as the mass parameter $M$ appearing the Klein-Gordon equation in 
an AdS$_5$ background, but is related to it by the $D=5$ case of the relation given in the introduction, i.e. $M^2 = M_c^2 + m^2$ (in units for which $\hbar=1$) 
where $M_c^2= -15/4$ (for $D=5$). At the level of classical particle mechanics, a particle of zero mass has a null worldline, and this corresponds to $m=0$ even in AdS. 

If  the gauge group were $U(1)\times SU(2)$ then $G(t)$ would be replaced by the composition of an $SU(2)$ transformation with a $U(1)$ transformation. Since  $\pi_1\left(SU(2)\right)$ is trivial we would have been able to focus exclusively on the $U(1)$ gauge invariance by setting  $G(t)= g(t) \bI_2$ for $g(t)\in U(1)$, in which case the global gauge anomaly is of the simpler $U(1)$ type discussed in  \cite{Elitzur:1985xj}. The maps $g(t)$ from a closed worldline to $U(1)$ fall into the integer homotopy classes of $\pi_1\left(U(1)\right)=\bZ$, and we may choose
$g_n(t) = \exp[-2n\pi it]$ as their representatives. In this case
\begin{equation}
G(t) = e^{-2n\pi i t} \bI_1 \quad \Rightarrow \quad  \tr_\mathbb{R}(i\dot GG^{-1}) = 4n\pi t\, , 
\end{equation}
and using this in  (\ref{deltaSm}) we deduce that 
\begin{equation}
e^{iS_m} \to  \left[e^{(2mR) 2\pi i}\right]^n e^{iS_m}\, , 
\end{equation}
and hence that $U(1)$ gauge invariance of the path integral requires the quantization condition 
\begin{equation}
2mR \in \bZ\, . 
\end{equation}
This quantization condition  is  implicit in the results of  \cite{Claus:1999jj}, but it is weaker than the quantization condition $mR\in \bZ$ required for $U(2)$ gauge invariance of the quantum path integral. The stronger quantization condition is also needed for the  total derivative term proportional to $mR$ in the key identity (\ref{key-identity}) to be an exact differential, as we have already remarked.

%%%%%%%%%%%%%%%%%%%%%%%%%%%%%%%%%%%%
%%%%%%%%%%%%%%%%%%%%%%%%%%%%%%%%%%%
%%%%%%%%%%%%%%%%%%%%%%%%%%%%%%%%%%
\section{The superparticle}\label{superparticlesect}

The starting point  in  \cite{Arvanitakis:2016vnp} for the construction of a supertwistor action for the superparticle in AdS$_D$ was the following action:
\begin{equation}\label{A1}
S=\int dt\left\{ \frac12\tr_{\mathbb R}\left[(\dot{\mathbb X}+\Theta_i^\dagger\dot\Theta^i-\dot\Theta_i^\dagger\Theta^i)\mathbb P\right]+\dot zp_z- \frac{1}{2}e (p^2 +\Delta^2) \right\}\,,
\end{equation}
where 
\begin{equation}
\Delta^2=p_z^2+\left(\frac{mR}z\right)^2\,, 
\end{equation}
and $\Theta^i$ is an  ${\cal N}$-plet ($i=1,\dots, {\cal N}$) of  $\bK$-valued matrices of {\it anticommuting} variables, acted upon from the left by the R-symmetry group $O(\mathcal N;\mathbb K)$ and on the right by $Sl(2;\mathbb K)$.  If the anticommuting variables are omitted then we recover
the action,  in Poincar\'e-patch coordinates, for a spin-zero particle of mass $m$ in AdS$_D$. 

The effect of the anticommuting variables is to  enlarge the Poincar\'e invariance  on Mink$_d$ slices to a super-Poincar\'e invariance. The above action is therefore invariant {\it by construction} under the action of a super-Poincar\'e group on the variables $\{\bX,\bP,\Theta^i\}$, under which the variables $\{z,p_z\}$ are inert. There is also a linearly 
realized $O({\cal N};\bK)$ R-symmetry, which acts only on the anticommuting variables, and a  scale invariance with respect to which $\Theta^i$ has dimension $-\tfrac{1}{2}$ if we assign dimension $-1$ to the AdS coordinates $\bX$ and $z$.

The motivation for this action comes from the observation of \cite{Mezincescu:2014zba} that a massive superparticle in a Mink$_d$ background has additional ``hidden'' supersymmetries. As AdS$_D$ is conformal to Mink$_D$, we may expect the action (\ref{A1}) to have additional ``hidden'' supersymmetries for $m=0$, and this is indeed the case. In fact, for $m=0$ the 
following constants of motion are Noether charges for an $OSp(\mathcal N|4;\mathbb K)$ invariance: 
\begin{eqnarray}\label{Noetherxpth}
\mathbb P &=&\mathbb P\,, \nonumber\\
\mathbb Q^i &=&\Theta^i\mathbb P\,, \nonumber\\
\mathbb D &=&\mathbb P(\mathbb X+\Theta^\dagger_k\Theta^k)+zp_z\mathbb I_2 \,, \qquad 
\mathbb R^i_{\phantom ij}=\Theta^i\mathbb P\Theta^\dagger_j\, ,\nonumber\\
\mathbb S^i&=&\Theta^i\left[\mathbb P\left(\mathbb X+\Theta^\dagger_j\Theta^j\right)+ zp_z\right]\,,\nonumber \\
\mathbb K &=&-(\mathbb X-\Theta^\dagger_k\Theta^k)\mathbb P(\mathbb X+\Theta^\dagger_l\Theta^l)-2zp_z\mathbb X+z^2\tilde{\mathbb P}\,.
\end{eqnarray}
These are constants of motion irrespective of whether $m$ is zero or non-zero but  the PBs of the $\bS^i$ charges close on $\bK$ (the Noether charge) to yield the expected algebra of $OSp(\mathcal N|4;\mathbb K)$ only if  $p^2+p_z^2=0$, which is the mass-shell constraint for $m=0$.

We conclude from this result that for $m=0$  the action (\ref{A1}) is an action for the massless superparticle in AdS$_D$. One should appreciate here that
this action is {\it much simpler} than the standard  one for a massless superparticle in an AdS background, because that action has a hidden fermionic gauge 
invariance \cite{Siegel:1983hh} that is generally called ``kappa-symmetry''. The results of \cite{Mezincescu:2014zba} for the Minkowski background case strongly suggest 
that  the action (\ref{A1}) is, at least for $m=0$, a gauge-fixed version of the kappa-symmetric action. We make no attempt here to verify this as none of the results to follow 
depend on its validity. 

For $m\ne0$ the PBs of the $\bS^i$ charges close on  $\bK' = \bK - (mR/\Delta)^2\tilde\bP\,$ but the PB of $\bK'$ with $\bS^i$ is non-zero, so one is led to a superalgebra with more generators than that of $OSp(\mathcal N|4;\mathbb K)$; we suspect that it is infinite dimensional (which would not be not surprising for a free particle).  We may attempt to rectify this problem  by modifying $\bS^i$ in addition to $\bK$. Although there is no such modification that resolves the problem for the general case, we may replace $\bS^i$ and $\bK$ in  the $D=5$ (complex)  case  by 
\begin{equation}\label{checkS}
\breve\bS^i = \bS^i -  i(mR)\Theta^i\, , \qquad \breve\bK = \bK - 2i(mR) \Theta_i^\dagger\Theta^i\, . 
\end{equation}
These are again constants of the motion that generate symmetries of the action, and the PBs of the $\breve\bS^i$ charges close on $\breve\bK$, 
as a consequence of the mass-shell constraint $p^2+\Delta^2=0$. The factors of $i$ are crucial to this result, and hence to existence of Noether charges spanning 
the algebra of $OSp(\mathcal N|4;\mathbb C) \cong U(2,2|N)$ for arbitrary $m$.

%%%%%%%%%%%%%%%%%%%%%%%%%%
%%%%%%%%%%%%%%%%%%%%%%%%%%
\subsection{Supertwistor formulation}

The construction of a two-supertwistor form of the action again starts by setting $\bP= \mp\bU\bU^\dagger$. The subsequent steps for $m=0$ 
were explained in \cite{Arvanitakis:2016vnp}; they involve the introduction of the new variables 
\begin{equation}\label{defss}
\bW=\pm\left(\bX - \Theta_i^\dagger\Theta^i\right)\bU -zp_z\bV^\dagger\, , \qquad 
\Xi^i=\Theta^i\mathbb U\, , 
\end{equation}
which satisfy an identity $\bG\equiv0$ for 
\begin{equation}\label{susyG}
\bG = \mathbb U^\dagger\mathbb W-\mathbb W^\dagger\mathbb U\pm2\Xi^\dagger_i\Xi^i\, . 
\end{equation}
This identity becomes a constraint in the action, imposed by an anti-hermitian Lagrange multipler $\bL$. This action is 
\begin{equation}\label{actions}
S=\int dt\,\tr_{\mathbb R}\big\{\dot{\mathbb U}\mathbb W^\dagger\mp\Xi_i^\dagger\dot\Xi^i-\mathbb L\mathbb G\big\}\, .
\end{equation}
The anticommuting variables are now scalars (with respect to the Mink$_d$ Lorentz group) appearing in the ${\cal N}$
$\bK$-valued matrices $\Xi^i$ which are now acted upon from the left by the R-symmetry group $O(\mathcal N;\mathbb K)$ 
and on the right by the $O(2;\mathbb K)$ gauge group, for which $\bG$ is the generator.  

Collectively, the new canonical variables are the   components of a two-supertwistor, i.e. an $O(2;\bK)$ doublet of spinors of  an $OSp(\mathcal N|4;\mathbb K)$ symmetry supergroup, whose generators are gauge-invariant supertwistor bilinears:
\begin{eqnarray}\label{OSP}
\bP  &=& \mp \bU\bU^\dagger\, , \nonumber \\
\bQ^i  &=&  \mp\Xi^i\mathbb U^\dagger\, , \nonumber \\
\bD &=&  -\mathbb U\mathbb W^\dagger\, , \qquad \bR^i{}_j = \mp\Xi^i\Xi^\dagger_j\, ,  \nonumber \\
\bS^i &=&  -\Xi^i\mathbb W^\dagger\, , \nonumber \\
\bK &=& = \pm\mathbb W\mathbb W^\dagger \, .
\end{eqnarray}
By using the relations (\ref{defss}) to rewrite these  two-supertwistor bilinears in terms of the variables $\{\bX,\bP,\Theta^i\}$ we confirm 
that they are indeed the AdS superisometry generators (\ref{Noetherxpth}) {\it provided that $m=0$}.  We could not have hoped for more than this because, as explained above, the action from which we started is only invariant under the AdS superisometries when $m=0$, {\it unless} $D=5$, 
which requires a separate analysis that we now present. 

For $D=5$ and $m\ne0$ we saw in section \ref{sec:D=5} that we should rewrite the (super)twistor action in terms of the new 
matrix variable 
\begin{equation}
\breve\bW= \bW - imR\, \bV^\dagger \, . 
\end{equation}
Omitting a total time derivative from the Lagrangian, the action (\ref{actions}) becomes 
\begin{equation}\label{check-actions}
S=\int dt\,\tr_{\mathbb R}\big\{\dot{\mathbb U}\breve{\mathbb W}^\dagger\mp\Xi_i^\dagger\dot\Xi^i-\mathbb L\mathbb G\big\}\, , 
\end{equation}
where $\bG$, when written in terms of $\breve\bW$, becomes
\begin{equation}
\bG = \mathbb U^\dagger\breve{\mathbb W}-\breve{\mathbb W}^\dagger\mathbb U\pm2\Xi^\dagger_i\Xi^i + 2imR\,  \bI_2\, . 
\end{equation}
As we saw in section \ref{sec:D=5} for the bosonic particle, the only change is a constant term proportional to $mR$ in the 
$U(1)$ constraint. 

The $OSp(\mathcal N|4;\mathbb C)$ generators are now formally the same as those in (\ref{OSP}) but with $\breve\bW$ in place of 
$\bW$. This adds a constant term to the $\bD$ generator but, for reasons explained in section \ref{sec:D=5}, this is just equivalent to 
the change in the $U(1)$ constraint. That leaves $\bS^i$ and $\bK$ which are replaced by
\begin{eqnarray}
\breve\bS^i = -\Xi^i\breve{\mathbb W}^\dagger  \, , \qquad \breve\bK =  \pm\breve{\mathbb W}\breve{\mathbb W}^\dagger\, . 
\end{eqnarray}
We now need to check that these definitions accord with those of (\ref{checkS}). First we consider
\begin{equation}
-\Xi^i\breve{\mathbb W}^\dagger = - \Theta^i \bU\bW^\dagger -imR\, \Theta^i = \bS^i -imR\, \Theta^i\, , 
\end{equation}
which is indeed the expression for $\breve\bS^i$ in (\ref{checkS}). Next we have
\begin{equation}\label{WW-prelim}
\pm\breve{\mathbb W}\breve{\mathbb W}^\dagger = \pm \bW\bW^\dagger 
\pm imR \, \bV^\dagger \left(\bU^\dagger\bW - \bW^\dagger\bU\right) \bV \pm (mR)^2\bV^\dagger\bV \, . 
\end{equation}
The $m=0$ equality  $\pm \bW\bW^\dagger=\bK$ relies on the $m=0$  mass-shell constraint; for $m\ne0$ we find, as in section \ref{sec:D=5}, that
\begin{equation}
\pm \bW\bW^\dagger = \bK \mp (mR)^2 \bV^\dagger\bV \, . 
\end{equation}
Using this, and the superparticle spin-shell constraint $\bG=0$, in (\ref{WW-prelim}) we have
\begin{equation}
\pm\breve{\mathbb W}\breve{\mathbb W}^\dagger =
\bK  - 2i mR\, \bV^\dagger \Xi_i^\dagger\Xi^i \bV  = \bK  - 2i mR \Theta_i^\dagger\Theta^i\,, 
\end{equation}
which is exactly $\breve\bK$ of (\ref{checkS}). 

We end with a comment on the sign of the fermion `kinetic' term in (\ref{actions}), or in (\ref{check-actions}). Recall that the upper sign is for 
positive energy and the lower sign for negative energy.  As first pointed out in \cite{Gauntlett:1990xq}, and further discussed in the context of twistor-type actions in  \cite{Mezincescu:2015apa}, this correlation is a required feature of {\it spacetime} supersymmetry in the context of the mechanics of particles,  or strings and branes, because it is needed for compatibility of spacetime supersymmetry with the existence of the negative energy states
that are  inevitable in  relativistic quantum mechanics.

%%%%%%%%%%%%%%%%%%%%
%%%%%%%%%%%%%%%%%%%%%%%%%
%%%%%%%%%%%%%%%%%%%%%%%
\section{Quantum spinning particle}

In the context of phase-space actions for particle mechanics with first-class phase-space constraints, the passage from classical  to quantum mechanics involves the  steps spelled out by Dirac.  First, the canonical variables are replaced by operators acting on some Hilbert space of quantum states, and their canonical Poisson  bracket relations are  replaced by canonical (anti)commutation relations. In this step the classical constraint functions become operators; ordering ambiguities  may arise but in the cases that we consider a choice of ordering exists such that the quantum constraints are also first-class (any remaining ambiguity is then intrinsic to the quantum theory). Next, ``physical'' states are 
taken to be those annihilated by the constraint operators.  In the case of reparametrization invariant phase-space actions with first-class constraints, as considered here,  these physical state conditions encode all physical properties of  the quantum particle.  

In our case, spin degrees of freedom arise from fermionic variables in the classical action, and a simplifying feature of the twistor formulation is that all these  variables are physical because there are no longer any fermionic constraints.  In the quantum theory these  variables become operators that act on a finite dimensional space of polarisation states.  Assuming a minimal realization of their canonical anticommutation relations, one can determine the dimension of this polarisation state space. 

This was done for the massless superparticle in \cite{Arvanitakis:2016vnp}. Here we perform a similar analysis for the massless $N$-extended spinning particle.  Our starting point will be the ``reduced'' action (\ref{simpleract}), which we recall here:
\begin{equation}\label{simpleractagain}
S = \int \! dt \left\{ \tfrac{1}{2}\tr_\mathbb{R}\left(\Omega\dot\bZ \bZ^\dagger + \tfrac{1}{2}\Psi_i\dot\Psi_i\right)  +  
\tr_\mathbb{R}\left[\bS \left(\bZ^\dagger \Omega \bZ + \Psi_i\Psi_i\right)\right] - \tfrac{1}{2}f_{ij}\tr_\mathbb{R} \left(\Psi_i\Psi_j \right) 
\right\}\, . 
\end{equation}
Classically,  this action governs the dynamics of a massless particle in AdS$_{4,5,7}$ with additional anticommuting variables.

%%%%%%%%%%%%%%%%%%%%%
%%%%%%%%%%%%%%%%%%%%%
\subsection{Canonical anticommutation relations}

We begin by choosing a convenient parametrization for the matrices $\Psi_i$ in terms of `real' anticommuting variables.  It will be sufficient to consider in detail the quaternionic case, for which we may write
\begin{equation}\label{param}
\Psi_i =\left( \begin{array}{cc}  \rho_i & \varsigma_i + {\bf i}\cdot{\bfpsi}_i \\  \varsigma_i - {\bf i}\cdot {\bfpsi}_i & -\rho_i \end{array}\right) \, , 
\end{equation}
and we replace ${\bf i}\cdot{\bfpsi}_i$ by $i\psi_i$ for $\bK=\bC$ and omit it for $\bK=\bR$.

As the matrices $\Psi_i$ transform by conjugation under $O(2;\bK)$,  which is the transverse rotation group in $D=3+{\rm dim}\, \bK$ 
(with an additional $U(1)$ factor for $D=5$) we should expect the components of $\Psi_i$ to transform (for each $i=1,\dots, N$) as a $(D-2)$ vector. To make this manifest, we 
define anticommuting variables $\{\vartheta^I; I=1,\dots,D-2\}$ such that (for $D=7$)
\begin{equation}
\vartheta_i^1 = \rho_i \, , \quad \vartheta_i^2 = \varsigma_i\, , \quad {\bfvartheta}_i = {\bfpsi}_i\, . 
\end{equation} 
We find that 
\begin{equation}
\frac{1}{4}\tr_\mathbb{R}(\Psi_i\dot\Psi_i)   = \frac{1}{2}\vartheta^I_i\dot\vartheta^I_i  \, , \qquad J_{ij} = \vartheta^I_i\vartheta^I_j \, . 
\end{equation}
These are manifestly ${\rm Spin}(5)$ invariant expressions, and from the first of them we may read off the canonical Poisson bracket relations, which become
the following canonical anticommutation relations for the corresponding hermitian operators $\hat\vartheta_i^I$ of the quantum theory:
\begin{equation}
\left\{\hat\vartheta_i^I,\hat\vartheta_j^J\right\} = \delta_{ij}\delta^{IJ}\, . 
\end{equation}
To include the real and complex cases we take $I=1, \dots, D-2$, where $D=4,5,7$.

%%%%%%%%%%%%%%%%%%%%%%%%%%%
\subsubsection{The global $SO(N)$ anomaly}

For $N>1$ real anticommuting  worldline variables, with standard canonical PB relations,  in the $N$-vector representation of a gauged $SO(N)$, the path integral  measure is ill-defined because of a global gauge anomaly \cite{Elitzur:1985xj}. This is easily verified for $N=3$ (the simplest case for which $SO(N)$ is non-abelian): the components of the anticommuting 3-vector become the Pauli matrices times a constant factor, so the polarisation states are 2-component $SU(2)$ spinors, but the $SO(3)$ generators that must annihilate these states are also Pauli matrices, so there are no physical states. 

This anomaly cancels if we have an {\it even} number of $N$-plets of real anticommuting variables \cite{Elitzur:1985xj}.  In the context of the ``reduced'' action (\ref{simpleractagain}) the number of such $N$-plets is $D-2$, as we have just seen, so the anomaly cancels for $D=4$ but {\it not} for $D=5,7$.  For $N=2$ and odd $D$, the anomaly can be cancelled by the inclusion of a WCS term, in principle, but this option is not available in the two-twistor form of the action, for the reasons explained in the Appendix.  

To summarize, the ``reduced'' action (\ref{simpleractagain}) will yield a consistent anomaly-free quantum theory for $D=4$,  and for  $D=5,7$ if $N=1$.
This conclusion is confirmed by the detailed analysis to follow. 

For $D=5,7$ and $N>1$ the reduced action leads to an inconsistent quantum theory because of the global $SO(N)$ gauge anomaly. 
This anomaly can be cancelled by  re-instating the redundant $N$-plet of anticommuting variables;  i.e. by reverting to the action (\ref{action_spinning_ads_twistor}). However, we expect this ``unreduced'' action  to  lead to  a reducible polarisation state space.  This is 
confirmed for our analysis below of the $N=2$ case; we do not attempt a detailed analysis of the $D=5,7$ cases for $N>2$.

%%%%%%%%%%%%%%%%%%%%%%%%%%%%
\subsubsection{Conformal invariance}

As the action (\ref{simpleractagain}) describes a particle of zero mass (classically) we should expect it to be invariant under the conformal 
isometries of AdS$_D$, as is the case for a Minkowski background \cite{Siegel:1988ru}. Given this, and assuming that conformal invariance is preserved upon quantization, we should expect to find
quantum wave equations in AdS$_D$ that are conformally invariant.

While the two-twistor formulation linearizes invariance under the $Sp(4;\bK)$ group of AdS$_D$ isometries, we cannot expect it to do the same
for the conformal isometries. However, as explained in \cite{Arvanitakis:2016vnp}, the larger conformal isometry group is also linearly realized for $D=4$. We should therefore expect to find  that conformal invariance  is preserved by the quantum theory at least in this case.

%%%%%%%%%%%%%%%%%%%%%%%
%%%%%%%%%%%%%%%%%%%%%%%%%%
\subsection{$N=1$}

We first consider the quantum theory for $N=1$.  In terms of the operators
\begin{equation}
\gamma^I = \sqrt{2} \, \hat\vartheta^I\, , 
\end{equation}
the canonical anticommutation relations are
\begin{equation}
\left\{\gamma^I,\gamma^J  \right\} = 2\delta^{IJ}\,  \qquad (I,J = 1,2,\dots D-2)\, . 
\end{equation}
These relations can be realized by Dirac matrices for $\bE^{D-2}$. Let us consider $D=4,5,7$ in turn:
\begin{itemize}

\item $D=4$. In this case we may choose $\gamma^1= \sigma_1$ and $\gamma^2= \sigma_3$. The state space is a 2-dimensional real vector space, so 
there are two polarisation states. This is what we should expect from a massless Dirac equation for a minimal spinor field in AdS$_4$; since the equation is conformally invariant and AdS$_4$
is conformal to Mink$_4$, the number of linearly independent polarisation states of the particle should be the same in AdS$_4$ as in Mink$_4$. 

\item $D=5$. In this case we may choose to realize the anticommutation relations in terms of the three $2\times2$ Pauli matrices, but these are complex so the state space is now a 2-dimensional 
complex vector space. This is equivalent to a 4-dimensional real vector space, so the number of independent polarisation states is now $4$. This is exactly what we should expect of a massless Dirac equation for a minimal spinor in AdS$_5$. The minimal spinor has 4 complex, or 8 real, components but only half are propagating; this is a standard result for the Dirac equation in Mink$_5$ but 
AdS$_5$ is conformal to Mink$_5$. 

\item $D=7$. In this case we may choose three of the 5 Dirac matrices to be $\bfsigma\otimes \sigma_1$, where $\bfsigma$ is the triplet of Pauli matrices, and then choose the other two to be 
$\bI_2\otimes\sigma_2$ and  $\bI_2\otimes\sigma_3$. This gives a  realization in terms of $4\times4$ complex matrices, which is equivalent to a realization in terms of $8\times8$ real matrices, 
so the state space is $8$-dimensional. Again, this is exactly what we should expect of a massless Dirac equation for a minimal spinor in AdS$_7$; in this case the minimal spinor has 16 real 
components, implying $8$ independent propagating modes. 

\end{itemize}
Notice that the polarisation state space has dimension $2\, {\rm dim}\, \bK$ for $D=3+ {\rm dim}\, \bK$. This is consistent with our expectation  
that the action (\ref{simpleractagain}) describes a massless particle, with spin $\tfrac{1}{2}$ (or the higher-dimensional equivalent) when $N=1$.

%%%%%%%%%%%%%%%%%%%%%%%%
%%%%%%%%%%%%%%%%%%%%%%
\subsection{$N=2$}

For $N=2$ we may replace the $(D-2)$ pairs of `real' anticommuting variables $\vartheta_i^I$ by the complex anticommuting variables
\begin{equation} 
\chi_I= \frac{1}{\sqrt{2}} \left(\vartheta_1^I + i\vartheta_2^I\right)\,  \qquad \left[\ \Rightarrow\ \bar\chi_I = \frac{1}{\sqrt{2}} \left(\vartheta_1^I - i\vartheta_2^I\right)\right]\, . \
\end{equation}
This gives us $J = i\chi_I\bar\chi_I$ for the classical $SO(2)$ constraint function. 

The canonical anticommutation relations of the quantum theory are now those of $(D-2)$ fermi oscillators: 
\begin{equation}
\left\{ \hat\chi_I, \hat\chi^\ddagger_J\right\} = \delta_{IJ} \, , 
\end{equation}
where $\ddagger$ indicates hermitian conjugation within the polarisation state space.  With standard operator ordering, the operator version 
of $J$ is\footnote{This step requires multiplication by $i$ in order to get an hermitian operator; recall that the factor of $i$
that is usually required  for `reality'  of  products of `real' anticommuting variables is absent in our conventions.}
\begin{equation}
\hat J =  \frac{1}{2}\sum_{I=1}^{D-2} \left[\hat\chi_I^\ddagger , \hat{\chi}_I\right] =  \sum_{I=1}^{D-2}  \hat\nu_I - \frac{D-2}{2} \, , 
\end{equation}
where the $\hat\nu_I$ are the $(D-2)$ fermi number operators  $\hat \nu_I = \hat\chi_I^\ddagger \hat\chi_I$ (no sum), with eigenvalues $\nu_I$ that are either $0$ or $1$. 
In a basis  for which the operators $\hat\nu_I$ are diagonal, the $SO(2)$ constraint becomes 
\begin{equation}\label{SO2}
 \sum_{I=1}^{D-2} \nu_I = \frac{D-2}{2}\,  \qquad ({\rm even}\ D)\, . 
 \end{equation}
 The restriction to even $D$ arises because there is no solution of this equation for odd $D$.  This is the well-known problem of a 
global $SO(2)\cong U(1)$ anomaly for an odd number of  fermi oscillators,  although the anomaly is actually the clash between $U(1)$ gauge invariance and discrete symmetries that are broken for a non-zero WCS term \cite{Elitzur:1985xj,Howe:1989vn}. The result of integrating out the  anticommuting variables, in the context of a path-integral quantization, 
is a WCS term with a coefficient  $c= \pm \tfrac{1}{2}$. If this is taken into account the  $SO(2)$ constraint in the form of (\ref{SO2}) is modified to 
\begin{equation}\label{SO2c}
 \sum_{I=1}^{D-2} \nu_I = \frac{D-2}{2} \pm \frac{1}{2}  \qquad \left({\rm odd}\ D\right), 
 \end{equation}
where one must choose one sign or the other. 

Although there is no fundamental reason to exclude the WCS term, we show in the Appendix that its inclusion creates a mismatch between the AdS isometry group and  the manifest $Sp(4;\bK)$ symmetry group of the action (\ref{simpleractagain}), so its inclusion in this context is problematic. However, the global $U(1)$ gauge anomaly can still be avoided by restoring the redundant anticommuting variables $\zeta'_i$ to the action, i.e. by reverting to the action (\ref{spinningact2}). In this case the  $SO(2)$ constraint becomes
\begin{equation}\label{SO2nu}
 \sum_{I=1}^{D-2} \nu_I + \nu_{\zeta'}= \frac{D-2}{2} - \frac{1}{2}\, , 
 \end{equation}
 where $\nu_{\zeta'}$ is the extra fermion occupation number, and the modification on the right hand side takes into account the zero point contribution of the additional fermi oscillator. After allowing for both possible values of $\nu_{\zeta'}$, this reduces to  the constraint (\ref{SO2c}) on $\nu_I$, but now {\it both signs are allowed}.
Thus, relative to the $c\ne 0$ resolution of the global $U(1)$ anomaly, the $\zeta_i'\ne0$ resolution leads to a doublet degeneracy in the polarisation state space. 

Let us now consider in turn the implications of these observations for  $D=4,5,7$. 

\begin{itemize}

\item $D=4$. In this case the $SO(2)$ constraint is $\hat\nu_1 + \hat\nu_2 = 1$, which has the two solutions 
\begin{equation}
(\nu_1,\nu_2) = \left\{(1,0) \, , \quad (0,1)\right\}\, . 
\end{equation}
There are therefore two polarisation states, which is what we should expect for a massless particle of spin $1$,  which 
is consistent with conformal invariance. 

\item $D=5$.  In this case we must use  (\ref{SO2c}), so the $SO(2)$ constraint is 
\begin{equation}
\nu_1 + \nu_2 + \nu_3 = \frac{3\pm 1}{2} \, . 
\end{equation}
For either choice of sign there are three solutions (either one state is empty and the other two full, or vice versa).  Apart from the doublet degeneracy if we take both sets
of three states, this is the expected result for a massless $2$-form or $3$-form gauge potential, but neither of these possibilities is consistent with conformal invariance. 

\item $D=7$. This case is similar to $D=5$. The $SO(2)$ constraint is 
\begin{equation}
\nu_1 + \nu_2 + \nu_3 +  \nu_4 + \nu_5 = \frac{5\pm 1}{2} \, . 
\end{equation}
For either choice of sign there are $10$ polarisation states, which is the number expected for massless  $3$-form or $4$-form gauge potential; again, neither of these possibilities is consistent with conformal invariance.  

\end{itemize}

%%%%%%%%%%%%%%%%%%%%%%%%
%%%%%%%%%%%%%%%%%%%%%%
\subsection{$N>2$ for $D=4$}

For $D=4$ we should expect the ``reduced'' action (\ref{simpleractagain}) to describe a massless particle of spin $N/2$ in AdS$_4$. 
We shall consider in detail only $N=3$ and $N=4$, from which the generalization to higher $N$ should be clear.  

\begin{itemize}

\item $N=3$. Recalling the choice $\gamma^1= \sigma_1$ and $\gamma^2= \sigma_3$ of 
Mink$_3$ Dirac matrices from the $N=1$ case, we have the following  $8\times8$ matrix realization of the $N=3$ anticommutation relations for $\gamma_i^I$:
\begin{eqnarray}\label{3Dirac}
\gamma_1^I  &=& \gamma^I\otimes \sigma_2 \otimes \bI_2 \nonumber \\
\gamma_2^I &=& \bI_2\otimes \gamma^I \otimes \sigma_2 \nonumber \\
\gamma_3^I &=& \sigma_2 \otimes \bI_2 \otimes\gamma^I
\end{eqnarray}
These matrices are not real but they are all pure imaginary, which implies that the $SO(3)$ generators are {\it real}:
\begin{eqnarray}
\hat J_{12} = - \frac{1}{2}\varepsilon_{IJ}\, \gamma^I\otimes \gamma^J \otimes i\sigma_2 \nonumber \\
\hat J_{23} = - \frac{1}{2}\varepsilon_{IJ}\,  i\sigma_2 \otimes \gamma^I\otimes \gamma^J \nonumber \\
\hat J_{31} =  \frac{1}{2}\varepsilon_{IJ}\,  \gamma^I \otimes i\sigma_2 \otimes \gamma^J\, . 
\end{eqnarray}
Because these matrices are real,  we may consistently suppose that the state space is real\footnote{This amounts to 
an assumption that the worldline time reversal invariance of the particle action becomes a discrete gauge invariance of the quantum theory \cite{Howe:1989vn}; see \cite{Arvanitakis:2016oyi} for a recent discussion.}.  Without the $SO(3)$ constraint we have an $8$-dimensional phase space, but 
now we must consider the effects of the constraints. 

The $4\times 4$ matrix $\varepsilon_{IJ} \gamma^I\otimes \gamma^J$ has eigenvalues $0,4$,  both doubly degenerate. The  zero-eigenvalue eigenvectors, 
i.e. zero modes, take the form $u\oplus i\sigma_2u$ for 2-component column vector $u$.  
A zero mode of $J_{12}$ is therefore $(u\oplus i\sigma_2 u) \oplus (v\oplus i\sigma_2 v)$ for two 2-component column vectors $(u,v)$.  
Requiring that this (real) $4$-vector also be  a zero mode of $J_{23}$ leads to the restriction $v=i\sigma_2 u$, so we now have a two-parameter space of  zero mode $8$-vectors of the form $u\oplus i\sigma_2 u \oplus i\sigma_2u\oplus (-u)$. These states are also annihilated by  
$J_{31}= -i[J_{12},J_{23}]$, so the $SO(3)$  constraints reduce the initial $8$-dimensional state space to a $2$-dimensional subspace.

\item $N=4$. We now have the following  {\it real} $16\times16$ matrix realization of the anticommutation relations:
\begin{eqnarray}\label{4Dirac}
\gamma_1^I  &=& \left[\gamma^I\otimes \sigma_2 \otimes \bI_2\right] \otimes \sigma_2 \nonumber \\
\gamma_2^I &=& \left[\bI_2\otimes \gamma^I \otimes \sigma_2\right]\otimes \sigma_2  \nonumber \\
\gamma_3^I &=& \left[\sigma_2 \otimes \bI_2 \otimes\gamma^I\right] \otimes \sigma_2 \nonumber \\
\gamma_4^I &=& \left[\, \bI_2 \otimes \, \bI_2 \otimes\,  \bI_2\right] \, \otimes \gamma^I\,  
\end{eqnarray}
where the brackets are included merely to emphasize that the enclosed $8\times 8$ matrices are those of (\ref{3Dirac}), and the $8\times8$
identity matrix. The $SO(4)$ generators are
\begin{equation}
\hat J_{ij} = \hat J_{ij}^{(N=3)} \otimes \bI_2 \qquad i,j =1,2,3
\end{equation}
which generate an $SO(3)$ subgroup, and 
\begin{eqnarray}
\hat J_{14} &=& - \frac{1}{2} \varepsilon_{IJ}\left[ \gamma^I \otimes i\sigma_2 \otimes \bI_2\right] \otimes \gamma^J  \nonumber \\
\hat J_{24} &=& - \frac{1}{2} \varepsilon_{IJ}\left[\bI_2 \otimes \gamma^I \otimes i\sigma_2\right]\otimes \gamma^J \nonumber \\
\hat J_{34} &=& - \frac{1}{2} \varepsilon_{IJ}\left[i\sigma_2 \otimes \bI_2 \otimes \gamma^I\right]\otimes \gamma^J\, . 
\end{eqnarray}
The $N=3$ results tell us that those $16$-vectors annihilated by the generators of the $SO(3)$ subgroup take the form
\begin{equation}
V= \left[u\oplus i\sigma_2 u \oplus i\sigma_2u\oplus (-u)\right]  \oplus \left[v\oplus i\sigma_2 v \oplus i\sigma_2v\oplus (-v)\right]\, , 
\end{equation}
where $u$ and $v$ are both two-component column vectors. These are annihilated by $\hat J_{14}$ if $v=i\sigma_2 u$, and then also by 
$\hat J_{24}$ and $\hat J_{34}$ as a consequence of $SO(3)$ invariance. The $SO(4)$ constraints therefore reduce the initial 
 $16$-dimensional state space to a $2$-dimensional subspace. 
 \end{itemize}
 Notice that the analysis for $N=4$ is essentially an iteration of that for $N=3$. Further iteration leads to the same conclusion 
 for $N>2$:  the physical subspace of polarisation states is two-dimensional, exactly as one expects for a massless particle of any spin in any conformally flat 4D spacetime, in particular AdS$_4$.

%%%%%%%%%%%%%%%%%%%%
%%%%%%%%%%%%%%%%%%%%%%%%%
%%%%%%%%%%%%%%%%%%%%%%%
\section{Discussion}

This paper is the continuation of a previous one \cite{Arvanitakis:2016vnp},  inspired by results of Claus et al. \cite{Claus:1999zh} and 
Cederwall \cite{Cederwall:2000km},  on a two-twistor formulation of  relativistic particle and superparticle mechanics  in a $D$-dimensional anti-de Sitter background,  for $D=4,5,7$. A novelty of  \cite{Arvanitakis:2016vnp} was the use of  $Sl(2;\bK)$-spinor notation where $\bK=\bR,\bC,\bH$ (the associative division algebras) to arrive at  a simple two-twistor  action for a massless particle in AdS$_D$, for $D=3+ {\rm dim}\, \bK$,   with a manifest $Sp(4;\bK)$ invariance.   For these spacetime dimensions the AdS isometry group is $Sp(4;\bK)$ (for an appropriate definition \cite{Sudbery,Chung:1987in}) and we verified that the $Sp(4;\bK)$ Noether charges  coincide with those associated to invariance  under AdS isometries. 

The starting point of the construction  is the observation that AdS$_D$ can be foliated by Mink$_d$ hypersurfaces, where $d=D-1$.  If the action 
for a massive particle in AdS$_D$ is expressed in the ``Poincar\'e-patch'' coordinates adapted to this foliation then its motion within any given hypersurface  is that of a particle in  Mink$_d$, with a mass $\Delta$ that is a particular constant of motion. The crucial step introduced in 
\cite{Arvanitakis:2016vnp} is to take $\Delta$ to be a new canonical variable and then, by means of a ``key identity'',  rewrite the action in terms of a  new set of canonical variables that include $\Delta$. At this point one may observe that the action for constant $\Delta$ {\it is} the action for a particle of mass  $\Delta$ in Mink$_d$,  which may be expressed in two-twistor form for $d=3,4,6$.  Subsequent elimination of $\Delta$ by means of the mass-shell constraint leads to  an action with manifest $Sp(4;\bK)$ invariance, and a local $O(2;\bK)$ gauge invariance associated to ``spin-shell'' constraints. 

The AdS isometry group is essentially $Sp(4;\bK)$ (it contains an additional $U(1)$ factor for $\bK=\bC$) and the canonical variables 
constitute a pair of  4-plets of this group. However, the  new action is independent of the mass $m$ originally assumed, and the  Noether charges of the manifest $Sp(4;\bK)$ invariance coincide with those of the  AdS isometry group  only if $m=0$, so only in this case do we have a manifestly linear realization of the AdS isometry group, with canonical variables that we can identify as two-twistor variables. 

This result raises the following question: at which point in the construction does the restriction to zero mass arise? 
The answer is not entirely clear to us but there is an obvious potential problem with the choice of Poincar\'e-patch coordinates because
these coordinates do not cover the whole of AdS. They cover only half of AdS (if we restrict to AdS rather than its covering space) with the 
two halves separated by a Killing horizon.  Massive particle  geodesics are curves that pass through both halves, so there is a global issue here for non-zero mass  that we passed over.  At the computational level,  one may observe  that  the ``key identity'' of (\ref{key-identity}) involves a total time derivative of $mR \varphi$, where (as explained in \cite{Arvanitakis:2016vnp}) $\varphi$ is an angular parameter on timelike geodesics. Strictly speaking, this term is not a  total time derivative because $d\varphi$ is not an exact 1-form. 

The construction of \cite{Claus:1999zh}  also uses a metric adapted to the foliation of AdS by Minkowski hypersurfaces, but it then proceeds
differently in a way that is specific to  AdS$_5$, in which case $O(2;\bC)\cong U(2)$, and the mass appears in the final result (in 
the combination $mR$, where $R$ is the AdS radius) as a contribution to the $U(1)$ charge. For $m=0$ the action agrees with the 
AdS$_5$ case of that found in  \cite{Arvanitakis:2016vnp}, and although it differs for $m\ne0$  there is a complex redefinition of the two-twistor 
variables of the $m=0$ action for an AdS$_5$ background  that changes only the $U(1)$ constraint, such that the result of \cite{Claus:1999zh}  is recovered. No such  redefinition is possible for AdS$_4$ or AdS$_7$, so only the massless particle has a two-twistor formulation in these cases, as originally shown by a very different method in \cite{Cederwall:2000km}. 

Even for AdS$_5$,  the two-twistor formulation of particle mechanics is incompatible with an {\it arbitrary} mass when we pass to the quantum theory.
We have shown that the absence of a global $U(2)$ gauge anomaly requires the quantization condition $mR\in \bZ$ (which also ensures that the 
ambiguity in the action due to the  $mRd\varphi$ term we neglected in our application of the ``key identity'' does not lead to an ambiguity in the 
path integral).  This $U(2)$ anomaly is similar to the  global $U(1)$ gauge anomaly in  quantum mechanics with anticommuting  variables discussed in Elitzur et al. \cite{Elitzur:1985xj} because it is essentially an anomaly in a  $U(1)$ subgroup of $U(2)$, but because $U(2)$ is a $\bZ_2$ quotient of $U(1)\times SU(2)$, the  quantization condition for $U(2)$ is stronger by a factor of $2$ than it would be for the $U(1)$ factor of $U(1)\times SU(2)$. 

However, the main new results of this paper are contained in an extension of the construction described above to the 
$N$-extended ``spinning particle'', which  generalizes the zero-spin  particle action to one that has $N$ local worldline supersymmetries. 
For a four-dimensional Minkowski background, this action is known to describe,  upon quantization,  a particle of  spin $N/2$, \cite{Brink:1976sz,Gershun:1979fb,Howe:1988ft}. Although the background spacetime is restricted to be conformally  flat for $N>2$,  this still allows the choice of  an AdS background \cite{Kuzenko:1995mg,Bastianelli:2014lia}.  

A crucial input is the two-twistor action  for the  {\it massive}  $N$-extended spinning particle in a Mink$_d$ background for $d=3,4,6$  \cite{Mezincescu:2015apa}.  The output is then a two-twistor action for  a {\it massless}  $N$-extended spinning particle in an AdS$_D$ background. For a massless particle one may expect an $N$-plet of anticommuting variables to be ``redundant'', as happens in the $m\to0$ limit of the standard  $N$-extended spinning particle action, and this is indeed the case. This has implications for the quantum theory because redundant anticommuting variables imply a degeneracy of polarisation states beyond that necessary for a massless particle of definite spin.  Omitting such redundant variables leaves us with a ``reduced'' action, and we have verified that  the $N=1$ version of it  leads to a polarisation  state space of the expected dimension  for a spin-$\tfrac{1}{2}$ particle in AdS$_D$. 

As mentioned above, a discrepancy for non-zero mass between the AdS Noether charges and those of the manifest $Sp(4;\bC)$ invariance arising from our construction of the two-twistor form of the spin-zero particle in AdS$_5$ can be eliminated by a change of variables that yields the action of \cite{Claus:1999zh}. For $N\ge1$ it is not possible to eliminate this discrepancy in the same way, even for  AdS$_5$ (and a similar problem arises for the massless $N=2$ spinning particle if a worldline Chern-Simons (WCS) term is included,  as we show in the Appendix).  Curiously, no similar difficulty arises for the superparticle, for any number of spacetime supersymmetries; we have shown that the manifest OSp invariance of the 
supertwistor action corresponds precisely with the expected AdS superisometries, even for the massive superparticle in AdS$_5$.

One advantage of the twistor formulation of the spinning particle  is that the anticommuting variables are all physical in the sense that they are
not subject to gauge transformations (in contrast to the standard formulation in which they are subject to local worldline supersymmetry transformations).  This simplifies an investigation into those aspects of the quantum theory that arise from the presence of anticommuting variables 
in the classical theory. Such an investigation was carried out in \cite{Arvanitakis:2016vnp} for the massless superparticle. Here we have presented
the results of a similar investigation for the massless $N$-extended spinning particle; an important consideration in this case is the possibility
of a global $SO(N)$ gauge anomaly of the type analysed in detail in  \cite{Elitzur:1985xj}. 

The implications of this quantum anomaly for the $N=2$ spinning particle in a Minkowski background were investigated in \cite{Howe:1989vn} (and there are also implications for the massless superparticle \cite{Arvanitakis:2016oyi}). The anomaly is really a clash between  the $SO(2)\cong U(1)$ gauge invariance and discrete symmetries that would be broken by a WCS term. If one demands preservation of the $U(1)$ gauge invariance then each worldline fermi-oscillator makes a contribution to the $U(1)$ charge that is equivalent to a contribution of $\pm 1/2$ to the coefficient of an effective WCS term. It is therefore impossible to maintain both $U(1)$ gauge invariance and a zero WCS term if the number of relevant  fermi oscilators is odd.  For the ``reduced'' $N=2$ two-twistor action this number is even when $D=4$ but odd for $D=5,7$. As we are  limited to zero WCS term in the twistor formulation (for reasons explained in the Appendix) we can avoid  the global gauge anomaly in these $D=5,7$ cases only by re-instating the redundant anticommuting variables (which amount to one fermi oscillator for $N=2$),  but this comes at the cost of a reducible polarisation state space. 

For $N>2$ there is a potential global $SO(N)$ anomaly \cite{Elitzur:1985xj}, and we have determined its implications for the quantum mechanics  of  the $N>2$ spinning particle.   As for $N=2$ case, the ``reduced'' two-twistor action for $N>2$ leads to an inconsistent quantum theory for $D=5,7$, which can again be remedied by the re-instatement of a redundant $N$-plet of anticommuting variables but, again, at the likely cost of a reducible polarisation state space. 

There are no global gauge anomalies for $N=1$, however.  In particular, for AdS$_5$ it should be  possible to extend the $N=0$ quantum results of \cite{Claus:1999jj} to $N=1$.   It should also be possible to make contact with the work of Adamo et al.  on a twistor formulation of free field equations  in AdS$_5$ \cite{Adamo:2016rtr}, as well as other related work  \cite{Uvarov:2016slb} and perhaps other approaches to particle mechanics in AdS backgrounds \cite{Heinze:2016lxs}. 

%%%%%%%%%%%%%%%%%%%%%%%%%%%%%%%
%%%%%%%%%%%%%%%%%%%%%%%%%%%%%%%%%
%%%%%%%%%%%%%%%%%%%%%%%%%%%%%%%%%
\appendix \section{N=2 and the WCS term}\label{sec:WCS}

The $N=2$ case of the spinning particle action (\ref{ADSbackground})  is special because the Lagrange multiplier $f_{ij} = \varepsilon_{ij} f$ for the $SO(2)\cong U(1)$ constraint is then a $U(1)$ gauge potential and we may  add to the action the  WCS term 
\begin{equation}
S_{WCS} = \hbar c\int\! f dt\, .  
\end{equation}
The factor of $\hbar$ multiplying the constant $c$ ensures that this term has the dimensions of action, and is needed anyway (unless one chooses units for which $\hbar=1)$ because  the WCS term should be considered as a possible local one-loop addition to the action. 

For $N=2$ we may write ${\cal J}_{ij} = \varepsilon_{ij} {\cal J}$, in which case 
\begin{equation}
\frac{1}{2} {\cal J}_{ij} {\cal J}_{ij}  = {\cal J}^2\, ,   \qquad {\cal J} = \frac{1}{2} \varepsilon^{ij} {\cal J}_{ij}\, . 
\end{equation}
Notice that this is consistent with our earlier definition of ${\cal J}^2$ in (\ref{scalarJ}) (and this is also true for $J^2$ if we write $J_{ij}= \varepsilon_{ij}J$). 
In this notation, the $f$-dependent term in the Lagrangian is now 
\begin{equation}
L_f = -f\left({\cal J}-\hbar c\right)\, . 
\end{equation}
Variation of $f$ therefore yields the modified $SO(2)$ constraint ${\cal J} = \hbar c$. This sets a bilinear in anticommuting variables equal to a real constant, which is an equation without solutions, but 
the  factor of $\hbar$ tells us that we should not be looking for classical solutions. 

The $U(1)$ gauge transformations associated to the $SO(2)$ constraint are 
\begin{equation}
(\xi_1 + i\xi_2) \to g(t)  (\xi_1 + i\xi_2) \qquad  {\it et\ cetera}\, , \qquad f \to f -ig^{-1}\dot g\, , 
\end{equation}
where $g(t)\in U(1)$ and ``{\it et cetera}'' stands for similar transformations of  all other fermionic variables. As discussed in section \ref{sec:D=5} in the 
context of the  other WCS that is possible for a (super)spin-zero (super)particle in AdS$_5$, the integral of $fdt$ is shifted by  $2\pi w$
for a ``large'' $U(1)$ gauge transformation for which $g(t)$ has winding number $w$, and this means that the path-integral is $U(1)$ gauge-invariant only if $c\in \bZ$, although this conclusion may be changed when ``fermionic'' variables are present. 

However, the task we set ourselves here is to determine how our results of section \ref{sec:AdStwist} are affected, in the $N=2$ case,  by the addition of a WCS term.  Our earlier derivation of the simplified constraints in Poincar\'e-patch coordinates  applies only for $c=0$, when specialized to $N=2$,  because  the {\it unmodified} $SO(2)$ constraint was used to simplify the other constraints. We should therefore expect some changes for $c\ne0$. 
For simplicity we now set $\hbar=1$.

%%%%%%%%%%%%%%%%%%%%%%%%%%%%%%%%%%%%%%%%
%%%%%%%%%%%%%%%%%%%%%%%%%%%%%%%%%%%%%%%%%
\subsection{Poincar\'e-patch coordinates redux}

It will suffice to focus on the effects of the WCS term on the supersymmetry generators ${\cal Q}_i$ as given by  (\ref{poincQ}). 
Previously, we simplified this expression for 
${\cal Q}_i$ using the $SO(N)$ constraint; we can now do the same again for $N=2$ but we must use the modified constraint, which can be writen as 
\begin{equation}
\lambda_i\cdot \lambda_j =  c\, \varepsilon_{ij} -\zeta_i\zeta_j -\xi_i\xi_j\, . 
\end{equation}
As before, this yields
\begin{equation}
{\cal Q}_i = \left(\frac{z}{R}\right)\tilde{\cal Q}_i  \, , \qquad \tilde{\cal Q}_i = p\cdot\lambda_i + \Xi_i\, , 
\end{equation}
but now
\begin{equation}\label{xic}
\Xi_i = p_z \zeta_i + \left(\frac{mR + \zeta_j\xi_j}{z}\right)\xi_i  + \frac{c}{z} \varepsilon_{ij} \zeta_j\, . 
\end{equation}
Previously, we used both the supersymetry and $SO(N)$ constraints to simplify the expression for ${\cal H}\equiv {\cal H}(0)$ in (\ref{Hzero})
but closure of the constraint algebra for $c\ne0$ eliminates the ambiguity in the Hamiltonian constraint function, which is necessarily
 ${\cal H}(1)$. Making this choice and then proceeding as before we find that 
\begin{equation}
2{\cal H}(1) = \left(\frac{z}{R}\right)^2\tilde{\cal H}\, , \qquad 2\tilde{\cal H} =  p^2+ \Delta^2\, , 
\end{equation}
where now
\begin{equation}\label{Deltac}
\Delta^2= p_z^2 + \left(\frac{mR + \zeta_k\xi_k}{z}\right)^2  + \left(\frac{c^2 + 2c(\zeta^2-\xi^2)}{z^2}\right)\, , 
\end{equation}
with
\begin{equation}
\zeta^2 \equiv  \frac{1}{2} \varepsilon_{ij} \zeta_i\zeta_j \, , \qquad \xi^2 \equiv  \frac{1}{2} \varepsilon_{ij} \xi_i\xi_j\, . 
\end{equation}
A Poisson bracket computation  shows, as before but now for $N=2$ with a WCS term, that
\begin{equation}\label{N2alg}
\left\{\tilde{\cal Q}_i, \tilde{\cal Q}_j\right\}_{PB} = 2 \tilde{\cal H}\delta_{ij}\,  , 
\end{equation}
and hence that the constraint functions generate a local $N=2$ worldline supersymmetry. However, the supersymmetry transformations are modified as a result of the modifications 
of the constraints. For any function $\phi$ of the phase-space variables, 
\begin{equation}
\delta_\epsilon \phi = \left\{\epsilon_i\tilde{\cal Q}_i, \phi \right\}_{PB}\, . 
\end{equation}
For the canonical variables this yields the $c$-dependent transformations
\begin{equation}\label{change}
\delta_\epsilon p_z =  \frac{mR}{z^2}\xi_i\epsilon_i -  \frac{(c-\xi^2)}{z^2}\varepsilon_{ij} \zeta_i\epsilon_j\, , 
\qquad \delta_\epsilon\zeta_i = p_z\epsilon_i - \frac{(c-\xi^2)}{z} \varepsilon_{ij} \epsilon_j\, , 
\end{equation}
along with the unchanged, and hence $c$-independent, transformations
\begin{eqnarray}\label{nochange}
\delta_\epsilon x^\mu &=& \lambda_i^\mu \epsilon_i\, , \qquad \delta_\epsilon z = \zeta_i \epsilon_i \, , \nonumber \\
\delta_\epsilon \lambda_i^\mu &=& \eta^{\mu\nu}p_\nu \epsilon_i\, \qquad \delta_\epsilon \xi_i = \left(\frac{mR+ \zeta_j \xi_j}{z}\right)\epsilon_i - z^{-1} \zeta_i\xi_j \epsilon_j\, . 
\end{eqnarray}
Using these results one may verify that $\Delta^2$, and hence $\tilde{\cal H}$, is still invariant under local supersymmetry, as required by the 
$N=2$ supersymmetry algebra (\ref{N2alg}). 

By analogy with the definition (\ref{xic}), which extends to $c\ne0$ the definition $\Xi_i$ of (\ref{defXi}), we can also extend to $c\ne0$ the definition $Z_i$ of (\ref{defZi}), which 
becomes
\begin{equation}
Z_i = p_z\xi_i - \left(\frac{mR + \zeta_j\xi_j}{z}\right)\zeta_i - \frac{c}{z} \varepsilon_{ij} \xi_j\, . 
\end{equation}
A PB calculation shows that we again have
\begin{equation}
\left\{Z_i,Z_j\right\}_{PB} = \Delta^2 \delta_{ij} \, , \qquad \left\{Z_i, \Xi_j\right\}_{PB} =0\, , 
\end{equation}
where $\Delta^2$ is now given by (\ref{Deltac}).  If we proceed as before to set
\begin{equation}
\Xi_i = \Delta\,  \xi'_i \, , \qquad Z_i = \Delta\,  \zeta'_i\, , 
\end{equation}
then we find, as before,  that the new primed anticommuting variables satisfy the simple PB relations 
\begin{equation}
\left\{\xi'_i,\xi'_j\right\}_{PB} = \delta_{ij} = \left\{\zeta'_i,\zeta'_j\right\}_{PB}\, , \qquad \left\{\xi'_i,\zeta'_j\right\}_{PB}=0\, . 
\end{equation}
The relation between the old anticommuting variables and the new, primed, ones is now significantly more complicated.  A detailed analysis, which we omit, 
shows that it is possible to rewrite the action in terms of the primed variables but the ``key identity'' relating the geometrical part of the action in the two sets of variables
allows us to take the next step towards a two-twistor action only if either $c=0$ or $m=0$. Since we are  interested in $c\ne0$
we shall proceed on the assumption that $m=0$.

%%%%%%%%%%%%%%%%%%%%%%%%%
%%%%%%%%%%%%%%%%%%%%%%%%%
\subsection{Zero mass}

For $m=0$ the expression (\ref{Deltac}) simplifies to 
\begin{equation}\label{Deltac0}
\Delta^2 = p_z^2 + \left(\frac{c + \zeta^2-\xi^2}{z}\right)^2\, , 
\end{equation}
which suggests that we define an angle $\varphi$ such that 
\begin{equation}\label{defangle2}
p_z = \Delta \cos\varphi\, , \qquad \left(\frac{c + \zeta^2-\xi^2}{z}\right)= \Delta\sin\varphi\, \, . 
\end{equation}
We then have 
\begin{eqnarray}
\xi'_i  =  \zeta_i \cos\varphi + \varepsilon_{ij}\zeta_j \sin\varphi \, , \qquad 
\zeta'_i = \xi_i \cos\varphi - \varepsilon_{ij}\xi_j \sin\varphi\, , 
\end{eqnarray}
and these relations imply
\begin{equation}
\xi^2 = (\zeta')^2\, , \qquad \zeta^2 = (\xi')^2\,  \quad \Rightarrow \quad \Delta^2 = p_z^2 + \left(\frac{c+ (\xi')^2 -(\zeta')^2}{z}\right)^2\, . 
\end{equation}
Proceeding as we did for the generic $N$ case in section \ref{sec:AdStwist} now leads to the identity
\begin{equation}\label{massless}
\dot z p_z + \frac{1}{2} \left( \zeta_i \dot\zeta_i + \xi_i\dot\xi_i \right)\equiv  - \frac{zp_z}{\Delta} \dot\Delta + \frac{1}{2} \left( \xi'_i \dot\xi'_i + \zeta'_i\dot\zeta'_i \right) 
+ \frac{d}{dt}\left( zp_z +c\varphi\right)\, .     
\end{equation}
Using this result we may proceed as before. After translating to $Sl(2;\bK)$-spinor notation we arrive at a Lagrangian 
that is formally identical to (\ref{bitwist-geom}) with local worldline supersymmetry constraints that are formally identical to those of (\ref{bitwist-cons}), but we are now restricted to $N=2$ 
and the $SO(2)$ constraint is ${\cal J}=c$ with\footnote{Recall that $\tilde\Lambda$ is the ``trace-reverse'' of the matrix $\Lambda$.}
\begin{equation}
{\cal J} = \frac{1}{4}  \tr_\mathbb{R} \left(\varepsilon^{ij} \tilde\Lambda_i \Lambda_j\right) + (\xi')^2 + (\zeta')^2\, . 
\end{equation}

We may now follow the steps of subsection \ref{sec:bitwistact}. We write $\bP= \mp \bU\bU^\dagger$ and substitute, solving the $\tilde{\cal H}=0$ constraint for $\Delta$ in terms of $\bU$. we then
solve the constraints ${\cal Q}_i=0$ for $\Lambda_i$ in terms of the new matrix variables $\Psi_i$, exactly as in (\ref{Lambda_def}) athough the $\Delta$ factor that appears in that expression 
for $\Lambda_i$ is now different.  This results in the new Lagrangian 
\begin{equation}
L= \tr_\mathbb{R}\left(\dot\bU\bW^\dagger\right) + \frac{1}{4}\tr_\mathbb{R} \left(\Psi_i \dot\Psi_i \right) + \frac{1}{2}\zeta'_i \dot\zeta'_i - f ({\cal J} -c)\, , 
\end{equation}
where  now
\begin{equation}
{\cal J} = \frac{1}{2} \varepsilon^{ij}  \tr_\mathbb{R}(\Psi_i\Psi_j) + \zeta'_i\zeta'_j\, ,  
\end{equation}
and $\bW$ is exactly as given in the incidence relation (\ref{W}), although now restricted to $N=2$. 
Because this incidence relation is unchanged,  the identity of (\ref{defbG}) still holds, with the same expression for $\bG$.
As before, this must be included as a constraint imposed by a Lagrange multiplier in  the action 
with $\bU$ and $\bW$  as independent variables.  After combining $\bU$ and $\bW$ into the two-twistor $\bZ$  we thus arrive at the Lagrangian 
\begin{equation}\label{spinningN=2c}
L_c= \frac{1}{2}\tr_\mathbb{R}(\Omega\dot\bZ \bZ^\dagger) + \frac{1}{4}\tr_\mathbb{R}(\Psi_i\dot\Psi_i) + \zeta'_i\dot\zeta'_i 
+ \tr_\mathbb{R}\left[\bS \left(\bZ^\dagger \Omega \bZ + \Psi_i\Psi_i\right)\right] - f ({\cal J}- c) \, .
\end{equation}

Of course, we could have deduced this result more directly by simply restricting  the two-twistor action of (\ref{action_spinning_ads_twistor}) to $N=2$ and then modifying 
the $SO(2)$ constraint to include the WCS term.  However, in order to check whether the manifest $Sp(4;\bK)$ invariance of the two-twistor Lagrangian $L_c$ is the invariance 
inherited from the isometry group of the AdS background, we need to rewrite the manifest $Sp(4;\bK)$ Noether charges  in terms of the variables of the original action. 
As this step involves details of the expression for $\Delta^2$, which differs from the expression used in  subsection \ref{sec:bitwistact},  it was necessary to arrive at 
(\ref{spinningN=2c}) by the longer route.

As we have seen in subsection \ref{sec:NC}, the only Noether charges of the manifest $Sp(4;\bK)$ invariance that are not guaranteed to arise from AdS isometries
are those contained in $\bW\bW^\dagger$. Our earlier discussion of this issue for the generic $N$ case implies, when restricted to $N=2$, that $\pm \bW\bW^\dagger = \bK$ 
when both $m=0$ and $c=0$, where $\bK$ is the AdS isometry matrix-charge of (\ref{bC}); in terms of the new, primed, variables this becomes
\begin{eqnarray}
\bK &=&  \frac{1}{2}\tilde\bP\left( \tr_\mathbb{R}(\bX\tilde\bX) + 2z^2 \right) -\bX\left( \tr_\mathbb{R}(\bX\bP) + zp_z\right) + 2\Lambda_i\xi'_i \frac{zp_z}{\Delta}  \nonumber \\
&& +\ \Lambda_i \tr_\mathbb{R}\left(\bX\tilde\Lambda_i\right)  - 2\Lambda_i \left(\frac{c+ (\xi')^2 -(\zeta')^2}{\Delta}\right) \varepsilon_{ij} \xi'_j \, . 
\end{eqnarray}
As we are now considering only $m=0$, any discrepancy between $\pm \bW\bW^\dagger$ and $\bK$
must be zero when $c=0$. This is confirmed by a direct calculation, which shows that 
\begin{equation}\label{discrepancy}
\pm \bW\bW^\dagger = \bK  \mp  \left[c- (\zeta')^2\right]^2 \bV^\dagger\bV\, . 
\end{equation}
The discrepancy between $\pm \bW\bW^\dagger$ and $\bK$ reduces to $\mp c \bV^\dagger\bV$ if we omit the $\zeta'_i$ variables on the grounds that they are redundant 
for $m=0$. Comparison with  (\ref{WWW2}) shows that the constant $c$ now replaces the constant $mR$; as in that case, the $c$-dependent term can be removed by a complex 
change of variables when $D=5$ but this leads to (\ref{discrep}) with $(mR)$ replaced by $c$, and since $\Psi_i\Psi_i$ is non-zero for $N=2$ there is still a discrepancy. 

We should stress that there is no fundamental obstruction to the inclusion of a WCS term in the action for the $N=2$ spinning particle, just as there is no fundamental  obstruction to the inclusion of a mass for $N>0$,  but both are obstructions to the existence of a two-twistor action,  even in the special case of AdS$_5$.

%%%%%%%%%%%%%%%%%%%%%%%%%%%%%%%%%%%%
\section*{Acknowledgements}

This work has been partially supported by the STFC consolidated grant ST/P000681/1.  A.S.A. also acknowledges support from Clare Hall College, Cambridge, and  the Cambridge Trust, and he  is grateful to Kostandinos Sfetsos and the Faculty of Physics at the University of Athens for their kind hospitality while part of this work was being undertaken; he was  also supported by the EPSRC programme grant  ``New Geometric Structures from String Theory'' (EP/K034456/1). A. B-G also acknowledges support from the SFTC grant 1495240.

%\bibliography{refsspin}

\begin{thebibliography}{1}

%\cite{Claus:1998mw}
\bibitem{Claus:1998mw}
  P.~Claus, R.~Kallosh, J.~Kumar, P.~K.~Townsend and A.~Van Proeyen,
  ``Conformal theory of M2, D3, M5 and D1-branes + D5-branes,''
  JHEP {\bf 9806} (1998) 004
  %doi:10.1088/1126-6708/1998/06/004
  [hep-th/9801206].
  %%CITATION = doi:10.1088/1126-6708/1998/06/004;%%

%\cite{Andrianopoli:1999kx}
\bibitem{Andrianopoli:1999kx}
  L.~Andrianopoli, M.~Derix, G.~W.~Gibbons, C.~Herdeiro, A.~Santambrogio and A.~Van Proeyen,
  ``Isometric embedding of BPS branes in flat spaces with two times,''
  Class.\ Quant.\ Grav.\  {\bf 17} (2000) 1875
  %doi:10.1088/0264-9381/17/9/301
  [hep-th/9912049].
  %%CITATION = doi:10.1088/0264-9381/17/9/301;%%


   %\cite{Penrose:1986ca}
\bibitem{Penrose:1986ca}
  R.~Penrose and W.~Rindler,
  ``Spinors and Space-time, Vol. 2'',  Oxford Univ. Press (1986). 
  %%CITATION = INSPIRE-238378;%%
  
  %\cite{Atiyah:2017erd}
\bibitem{Atiyah:2017erd}
  M.~Atiyah, M.~Dunajski and L.~Mason,
  ``Twistor theory at fifty: from contour integrals to twistor strings,''
  arXiv:1704.07464 [hep-th].
  %%CITATION = ARXIV:1704.07464;%%
  
  %\cite{Ferber:1977qx}
\bibitem{Ferber:1977qx}
  A.~Ferber,
  ``Supertwistors and Conformal Supersymmetry'', 
  Nucl.\ Phys.\ B {\bf 132} (1978) 55.
  %%CITATION = NUPHA,B132,55;%%
  
  %\cite{Shirafuji:1983zd}
\bibitem{Shirafuji:1983zd}
 T.~Shirafuji,
  ``Lagrangian Mechanics of Massless Particles With Spin'', 
  Prog.\ Theor.\ Phys.\  {\bf 70} (1983) 18.
 % doi:10.1143/PTP.70.18
  %%CITATION = doi:10.1143/PTP.70.18;%%


%\cite{Claus:1999zh}
\bibitem{Claus:1999zh}
  P.~Claus, J.~Rahmfeld and Y.~Zunger,
  ``A Simple particle action from a twistor parametrization of AdS(5),''
  Phys.\ Lett.\ B {\bf 466} (1999) 181
  %doi:10.1016/S0370-2693(99)01128-4
  [hep-th/9906118].
  %%CITATION = doi:10.1016/S0370-2693(99)01128-4;%%
  
   
  %\cite{Bengtsson:1987si}
\bibitem{Bengtsson:1987si}
  I.~Bengtsson and M.~Cederwall,
  ``Particles, Twistors and the Division Algebras,''
  Nucl.\ Phys.\ B {\bf 302} (1988) 81.
  %doi:10.1016/0550-3213(88)90667-0
  %%CITATION = doi:10.1016/0550-3213(88)90667-0;%%
  
   
  %\cite{Cederwall:1993xe}
\bibitem{Cederwall:1993xe}
  M.~Cederwall,
  ``Introduction to division algebras, sphere algebras and twistors,''
  hep-th/9310115.
  %%CITATION = HEP-TH/9310115;%%

  
   %\cite{Kugo:1982bn}
\bibitem{Kugo:1982bn}
  T.~Kugo and P.~K.~Townsend,
  ``Supersymmetry and the Division Algebras,''
  Nucl.\ Phys.\ B {\bf 221} (1983) 357.
  %doi:10.1016/0550-3213(83)90584-9
  %%CITATION = doi:10.1016/0550-3213(83)90584-9;%%

  
 %\cite{Sudbery}
\bibitem{Sudbery}
  A.~Sudbery,
  ``Division algebras, (pseudo)orthogonal groups and spinors'',
  J. Phys. {\bf A17} (1984) 939.
  
  %\cite{Manogue:1993ja}
\bibitem{Manogue:1993ja}
  C.~A.~Manogue and J.~Schray,
  ``Finite Lorentz transformations, automorphisms, and division algebras,''
  J.\ Math.\ Phys.\  {\bf 34} (1993) 3746,
 % doi:10.1063/1.530056
  [hep-th/9302044].
  %%CITATION = doi:10.1063/1.530056;%%
  
  %\cite{Viero}
  \bibitem{Viero}
  J.P. Viero, ``Octonionic presentation for the Lie group $Sl(2,\bO)$'', Journal of Algebra and its Applications {\bf 13} (2014) 1450017,
  %doi:10.1142/S0219498814500170
 [arXiv:1504.04065 [math.DG]].
 
 
   %\cite{Chung:1987in}
\bibitem{Chung:1987in}
  K.~W.~Chung and A.~Sudbery,
  ``Octonions and the Lorentz and Conformal Groups of Ten-dimensional Space-time,''
  Phys.\ Lett.\ B {\bf 198} (1987) 161.
 % doi:10.1016/0370-2693(87)91489-4
  %%CITATION = doi:10.1016/0370-2693(87)91489-4;%%
  
   %\cite{Howe:1992bv}
\bibitem{Howe:1992bv}
  P.~S.~Howe and P.~C.~West,
  ``The Conformal group, point particles and twistors,''
  Int.\ J.\ Mod.\ Phys.\ A {\bf 7} (1992) 6639.
  %doi:10.1142/S0217751X92003057
  %%CITATION = doi:10.1142/S0217751X92003057;%%

  
   
   %\cite{Cederwall:2000km}
\bibitem{Cederwall:2000km}
  M.~Cederwall,
  ``Geometric construction of AdS twistors'', 
  Phys.\ Lett.\ B {\bf 483} (2000) 257,  
%  doi:10.1016/S0370-2693(00)00552-9
  [hep-th/0002216]; 
  %%CITATION = doi:10.1016/S0370-2693(00)00552-9;%%
  %\cite{Cederwall:2004cf}
   %\cite{ Cederwall:2004cf}
%\bibitem{Cederwall:2004cf}
%M.~Cederwall,
  ``AdS twistors for higher spin theory,''
  AIP Conf.\ Proc.\  {\bf 767} (2005) 96, 
 % doi:10.1063/1.1923331
  [hep-th/0412222].
  %%CITATION = doi:10.1063/1.1923331;%%
  
  
  
  
  %\cite{Hawking:1973uf}
\bibitem{Hawking:1973uf}
  S.~W.~Hawking and G.~F.~R.~Ellis,
  ``The Large Scale Structure of Space-Time,'' CUP, 1982. 
  %doi:10.1017/CBO9780511524646
  %%CITATION = doi:10.1017/CBO9780511524646;%%

  %\cite{Breitenlohner:1982jf}
\bibitem{Breitenlohner:1982jf}
  P.~Breitenlohner and D.~Z.~Freedman,
  ``Stability in Gauged Extended Supergravity'',
  Annals Phys.\  {\bf 144} (1982) 249.
  %doi:10.1016/0003-4916(82)90116-6
  %%CITATION = doi:10.1016/0003-4916(82)90116-6;%%  
  
  %\cite{Mezincescu:1984ev}
\bibitem{Mezincescu:1984ev}
  L.~Mezincescu and P.~K.~Townsend,
  ``Stability at a Local Maximum in Higher Dimensional Anti-de Sitter Space and Applications to Supergravity'',
  Annals Phys.\  {\bf 160} (1985) 406.
%  doi:10.1016/0003-4916(85)90150-2
  %%CITATION = doi:10.1016/0003-4916(85)90150-2;%%


  %\cite{Arvanitakis:2016vnp}
\bibitem{Arvanitakis:2016vnp}
  A.~S.~Arvanitakis, A.~E.~Barns-Graham and P.~K.~Townsend,
  ``Anti de Sitter Particles and Manifest (Super)Isometries,''
  Phys.\ Rev.\ Lett.\  {\bf 118} (2017) no.14,  141601
 % doi:10.1103/PhysRevLett.118.141601
  [arXiv:1608.04380 [hep-th]].
  %%CITATION = doi:10.1103/PhysRevLett.118.141601;%%
  
   %\cite{Fedoruk:2004ru}
\bibitem{Fedoruk:2004ru}
  S.~Fedoruk and V.~G.~Zima,
  ``Bitwistor formulation of the spinning particle,''
  hep-th/0401064.
  %%CITATION = HEP-TH/0401064;%%
  
  %\cite{deAzcarraga:2005ky}
\bibitem{deAzcarraga:2005ky}
  J.~A.~de Azcarraga, A.~Frydryszak, J.~Lukierski and C.~Miquel-Espanya,
  ``Massive relativistic particle model with spin from free two-twistor dynamics and its quantization,''
  Phys.\ Rev.\ D {\bf 73} (2006) 105011
  %doi:10.1103/PhysRevD.73.105011
  [hep-th/0510161].
  %%CITATION = doi:10.1103/PhysRevD.73.105011;%%

 %\cite{Jiusi:2017xiy}
\bibitem{Jiusi:2017xiy}
  L.~Jiusi and V.~P.~Nair,
  ``Actions for particles and strings and Chern-Simons gravity,''
  Phys.\ Rev.\ D {\bf 96} (2017) no.6,  065019
  %doi:10.1103/PhysRevD.96.065019
  [arXiv:1706.05021 [hep-th]].  
  
  %\cite{Casalbuoni:1976tz}
\bibitem{Casalbuoni:1976tz}
  R.~Casalbuoni,
  ``The Classical Mechanics for Bose-Fermi Systems,''
  Nuovo Cim.\ A {\bf 33} (1976) 389.
%  doi:10.1007/BF02729860
  %%CITATION = doi:10.1007/BF02729860;%%
  
  %\cite{Brink:1981nb}
\bibitem{Brink:1981nb}
  L.~Brink and J.~H.~Schwarz,
  ``Quantum Superspace,''
  Phys.\ Lett.\  {\bf 100B} (1981) 310.
 % doi:10.1016/0370-2693(81)90093-9
  %%CITATION = doi:10.1016/0370-2693(81)90093-9;%%

  %\cite{Siegel:1983hh}
\bibitem{Siegel:1983hh}
  W.~Siegel,
  ``Hidden Local Supersymmetry in the Supersymmetric Particle Action,''
  Phys.\ Lett.\  {\bf 128B} (1983) 397.
  doi:10.1016/0370-2693(83)90924-3
  %%CITATION = doi:10.1016/0370-2693(83)90924-3;%%

  
  
 

 %\cite{Claus:1999jj}
\bibitem{Claus:1999jj}
  P.~Claus, R.~Kallosh and J.~Rahmfeld,
  ``BRST quantization of a particle in AdS(5),''
  Phys.\ Lett.\ B {\bf 462} (1999) 285
  %doi:10.1016/S0370-2693(99)00931-4
  [hep-th/9906195].
  %%CITATION = doi:10.1016/S0370-2693(99)00931-4;%%

  %\cite{Claus:1999xr}
\bibitem{Claus:1999xr}
  P.~Claus, M.~Gunaydin, R.~Kallosh, J.~Rahmfeld and Y.~Zunger,
  ``Supertwistors as quarks of SU(2, 2|4),''
  JHEP {\bf 9905} (1999) 019
  %doi:10.1088/1126-6708/1999/05/019
  [hep-th/9905112].
  %%CITATION = doi:10.1088/1126-6708/1999/05/019;%%



 %\cite{Brink:1976sz}
\bibitem{Brink:1976sz}
  L.~Brink, S.~Deser, B.~Zumino, P.~Di Vecchia and P.~S.~Howe,
  ``Local Supersymmetry for Spinning Particles,''
  Phys.\ Lett.\ B {\bf 64} (1976) 435
   Erratum: [Phys.\ Lett.\ B {\bf 68} (1977) 488].
 % doi:10.1016/0370-2693(76)90115-5
  %%CITATION = doi:10.1016/0370-2693(76)90115-5;%%
  
  %\cite{Brink:1976uf}
\bibitem{Brink:1976uf}
  L.~Brink, P.~Di Vecchia and P.~S.~Howe,
  ``A Lagrangian Formulation of the Classical and Quantum Dynamics of Spinning Particles,''
  Nucl.\ Phys.\ B {\bf 118} (1977) 76.
 % doi:10.1016/0550-3213(77)90364-9
  %%CITATION = doi:10.1016/0550-3213(77)90364-9;%%

  
    %\cite{Gershun:1979fb}
\bibitem{Gershun:1979fb}
  V.~D.~Gershun and V.~I.~Tkach,
  ``Classical And Quantum Dynamics Of Particles With Arbitrary Spin,''
  JETP Lett.\  {\bf 29} (1979) 288
   [Pisma Zh.\ Eksp.\ Teor.\ Fiz.\  {\bf 29} (1979) 320].
  %%CITATION = JTPLA,29,288;%%
  
 
%\cite{Howe:1988ft}
\bibitem{Howe:1988ft}
  P.~S.~Howe, S.~Penati, M.~Pernici and P.~K.~Townsend,
  ``Wave Equations for Arbitrary Spin From Quantization of the Extended Supersymmetric Spinning Particle,''
  Phys.\ Lett.\ B {\bf 215} (1988) 555.
  %%CITATION = PHLTA,B215,555;%%

  %\cite{Howe:1989vn}
\bibitem{Howe:1989vn}
  P.~S.~Howe, S.~Penati, M.~Pernici and P.~K.~Townsend,
  ``A Particle Mechanics Description of Antisymmetric Tensor Fields,''
  Class.\ Quant.\ Grav.\  {\bf 6} (1989) 1125.
  %%CITATION = CQGRD,6,1125;%%
  
  
   %\cite{Kuzenko:1995mg}
\bibitem{Kuzenko:1995mg}
  S.~M.~Kuzenko and Z.~V.~Yarevskaya,
  ``Conformal invariance, N extended supersymmetry and massless spinning particles in anti-de Sitter space,''
  Mod.\ Phys.\ Lett.\ A {\bf 11} (1996) 1653
  %doi:10.1142/S0217732396001648
  [hep-th/9512115].
  %%CITATION = doi:10.1142/S0217732396001648;%%

  
%\cite{Bastianelli:2014lia}
\bibitem{Bastianelli:2014lia}
  F.~Bastianelli, R.~Bonezzi, O.~Corradini and E.~Latini,
  ``Massive and massless higher spinning particles in odd dimensions,''
  JHEP {\bf 1409} (2014) 158
 % doi:10.1007/JHEP09(2014)158
  [arXiv:1407.4950 [hep-th]].
  %%CITATION = doi:10.1007/JHEP09(2014)158;%%
      
  
  % \cite{Mezincescu:2015apa}
\bibitem{Mezincescu:2015apa}
  L.~Mezincescu, A.~J.~Routh and P.~K.~Townsend,
  ``Twistors and the massive spinning particle,''
  J.\ Phys.\ A {\bf 49} (2016) no.2,  025401
  %doi:10.1088/1751-8113/49/2/025401
  [arXiv:1508.05350 [hep-th]].
  %%CITATION = doi:10.1088/1751-8113/49/2/025401;%%
  
   %\cite{Arvanitakis:2016wdn}
\bibitem{Arvanitakis:2016wdn}
  A.~S.~Arvanitakis, L.~Mezincescu and P.~K.~Townsend,
  ``Pauli-Lubanski, Supertwistors, and the Superspinning Particle,''
  JHEP {\bf 1706} (2017) 151
% doi:10.1007/JHEP06(2017)151
  [arXiv:1601.05294 [hep-th]].
  %%CITATION = doi:10.1007/JHEP06(2017)151;%%

  
   
  %\cite{Gilmore}
   \bibitem{Gilmore}
   R. Gilmore, ``Lie Groups, Lie Algebras and Some of Their Applications'', 
   Wiley-Interscience,  1974.
   
  
%\cite{Alslaken}
\bibitem{Aslaken}
H. Aslaken, Quaternionic determinants, The Mathematical Intelligencer 18 (1996) 57Ð65.



  %\cite{Schray:1994fc}
\bibitem{Schray:1994fc}
  J.~Schray,
  ``The General classical solution of the superparticle,''
  Class.\ Quant.\ Grav.\  {\bf 13} (1996) 27
  %doi:10.1088/0264-9381/13/1/004
  [hep-th/9407045].
  %%CITATION = doi:10.1088/0264-9381/13/1/004;%%


%  \cite{Zhang}
\bibitem{Zhang}
F. Zhang, ``Quaternions and Matrices of Quaternions'', Linear Algebra and its Applications 251 (1997) 21-57. 

%\cite{Routh:2015ifa}
\bibitem{Routh:2015ifa}
  A.~J.~Routh and P.~K.~Townsend,
  ``Twistor form of massive 6D superparticle,''
  J.\ Phys.\ A {\bf 49} (2016) no.2,  025402
  %doi:10.1088/1751-8113/49/2/025402
  [arXiv:1507.05218 [hep-th]].
  %%CITATION = doi:10.1088/1751-8113/49/2/025402;%%
  
     %\cite{Elitzur:1985xj}
\bibitem{Elitzur:1985xj}
  S.~Elitzur, Y.~Frishman, E.~Rabinovici and A.~Schwimmer,
  ``Origins of Global Anomalies in Quantum Mechanics,''
  Nucl.\ Phys.\ B {\bf 273} (1986) 93.
  %%CITATION = NUPHA,B273,93;%%
  
   %\cite{Mezincescu:2014zba}
\bibitem{Mezincescu:2014zba}
  L.~Mezincescu, A.~J.~Routh and P.~K.~Townsend,
  ``All Superparticles are BPS,''
  J.\ Phys.\ A {\bf 47} (2014) 175401
  %doi:10.1088/1751-8113/47/17/175401
  [arXiv:1401.5116 [hep-th]].
  %%CITATION = doi:10.1088/1751-8113/47/17/175401;%%
  



   %\cite{Gauntlett:1990xq}
\bibitem{Gauntlett:1990xq}
  J.~P.~Gauntlett, J.~Gomis and P.~K.~Townsend,
  ``Supersymmetry and the physical phase space formulation of spinning particles,''
  Phys.\ Lett.\ B {\bf 248} (1990) 288.
  %doi:10.1016/0370-2693(90)90294-G
  %%CITATION = doi:10.1016/0370-2693(90)90294-G;%%

  
    
 
  
   %\cite{Siegel:1988ru}
\bibitem{Siegel:1988ru}
  W.~Siegel,
  ``Conformal Invariance of Extended Spinning Particle Mechanics,''
  Int.\ J.\ Mod.\ Phys.\ A {\bf 3} (1988) 2713.
  %doi:10.1142/S0217751X88001132
  %%CITATION = doi:10.1142/S0217751X88001132;%%




  
    %\cite{Arvanitakis:2016oyi}
\bibitem{Arvanitakis:2016oyi}
  A.~S.~Arvanitakis, L.~Mezincescu and P.~K.~Townsend,
  ``Worldline CPT and massless supermultiplets,''
  Int.\ J.\ Mod.\ Phys.\ A {\bf 31} (2016) no.27,  1650152
  %doi:10.1142/S0217751X16501529
  [arXiv:1607.00526 [hep-th]].
  %%CITATION = doi:10.1142/S0217751X16501529;%%
  
    
 
 %\cite{Adamo:2016rtr}
\bibitem{Adamo:2016rtr}
  T.~Adamo, D.~Skinner and J.~Williams,
  ``Twistor methods for AdS$_{5}$,''
  JHEP {\bf 1608} (2016) 167
  %doi:10.1007/JHEP08(2016)167
  [arXiv:1607.03763 [hep-th]].
  %%CITATION = doi:10.1007/JHEP08(2016)167;%%  
  
  %\cite{Uvarov:2016slb}
\bibitem{Uvarov:2016slb}
  D.~V.~Uvarov,
  ``Ambitwistors, oscillators and massless fields on $AdS_5$,''
  Phys.\ Lett.\ B {\bf 762} (2016) 415
  %doi:10.1016/j.physletb.2016.09.065
  [arXiv:1607.05233 [hep-th]].
  %%CITATION = doi:10.1016/j.physletb.2016.09.065;%%
  
  %\cite{Heinze:2016lxs}
\bibitem{Heinze:2016lxs}
  M.~Heinze, G.~Jorjadze and L.~Megrelidze,
  ``Coset construction of AdS particle dynamics,''
  J.\ Math.\ Phys.\  {\bf 58} (2017) no.1,  012301
  %doi:10.1063/1.4974322
  [arXiv:1610.08212 [hep-th]].
  %%CITATION = doi:10.1063/1.4974322;%%
  
  


\end{thebibliography}
%\bibliographystyle{toine}

\providecommand{\href}[2]{#2}\begingroup\raggedright\endgroup

\end{document}